\documentclass[iop,twocolappendix]{emulateapj}

\usepackage{graphicx}
\usepackage{amsmath}
\usepackage{amssymb}
\usepackage{enumerate}
\usepackage[]{natbib}
\usepackage[usenames]{color} 
\usepackage{ulem}
\newcounter{address}
\newcommand{\kpc}{\ensuremath{\,\mathrm{kpc}}}
\newcommand{\kms}{\ensuremath{\,\mathrm{km\ s}^{-1}}}

\newcommand{\eq}[1]{\begin{align}#1\end{align}}

\newcommand{\PyGaia}{\texttt{{PyGaia}}}
\newcommand{\etal}{{et al.~}}

\def\apj{{Astroph. J.}}
\def\apjs{{Astroph. J. Suppl.}}
\def\aj{{Astron. J.}}
\def\mnras{{MNRAS}}

\shorttitle{Simulation velocity anisotropy parameter ($\beta$)}

\shortauthors{Loebman \etal}
\begin{document}

\title{Beta dips in the \textit{Gaia} era: simulation predictions of the Galactic velocity anisotropy parameter ($\beta$) for stellar halos}

\author{Sarah R.~Loebman\altaffilmark{1,2,3,4}}
\author{Monica Valluri\altaffilmark{1}}
\author{Kohei Hattori\altaffilmark{1}}
\author{Victor P.~Debattista\altaffilmark{5}}
\author{Eric F.~Bell\altaffilmark{1}}
\author{Greg Stinson\altaffilmark{6}}
\author{Charlotte R.~Christensen\altaffilmark{7}}
\author{Alyson Brooks\altaffilmark{8}}
\author{Thomas R.~Quinn\altaffilmark{9}}
\author{Fabio Governato\altaffilmark{9}}

\altaffiltext{1}{{Department of Astronomy, University of Michigan,
                 1085 S.~University Ave, Ann Arbor, MI 48109-1107, USA}}
\altaffiltext{2}{Michigan Society of Fellows}
\altaffiltext{3}{Hubble fellow}
\altaffiltext{4}{{Department of Physics, University of California, Davis,
                 1 Shields Ave, Davis, CA 95616, USA};        
		 {\tt srloebman@ucdavis.edu}}
\altaffiltext{5}{Jeremiah Horrocks Institute, University of Central Lancashire,
                 Preston PR1 2HE, UK}
\altaffiltext{6}{Max-Planck-Institut f\"ur Astronomie, Heidelberg, Germany}
\altaffiltext{7}{Physics Department, Grinnell College, Grinnell, IA, USA}
\altaffiltext{8}{Department of Physics \& Astronomy, 
                 Rutgers University, New Brunswick, NJ, USA}
\altaffiltext{9}{Astronomy  Department, University of  Washington, 
                 Seattle, WA, USA} 
\begin{abstract}
The velocity anisotropy parameter, $\beta$, is a measure of the kinematic state of orbits in the stellar halo which holds promise for constraining the merger history of the Milky Way (MW).
We determine global trends for $\beta$ as a function of radius from three suites of simulations, including accretion only and cosmological hydrodynamic simulations.
We find that both types of simulations are consistent and predict strong radial anisotropy ($<$$\beta$$>\sim0.7$) for Galactocentric radii greater than 10 kpc.
Previous observations of $\beta$ for the MW's stellar halo claim a detection of an isotropic or tangential ``dip'' at $r\sim20$~kpc.
Using $N$--body+SPH simulations, we investigate the temporal persistence, population origin, and severity of ``dips'' in $\beta$.
We find dips in the \textit{in situ} stellar halo are long-lived, while dips in the accreted stellar halo are short-lived and tied to the recent accretion of satellite material.
We also find that a major merger as early as $z\sim1$ can result in a present day low (isotropic to tangential) value of $\beta$ over a broad range of radii and angular expanse.
While all of these mechanisms are plausible drivers for the $\beta$ dip observed in the MW, in the simulations, each mechanism has a unique metallicity signature associated with it, implying that future spectroscopic surveys could distinguish between them.
Since an accurate knowledge of $\beta(r)$ is required for measuring the mass of the MW halo, we note significant transient dips in $\beta$ could cause an overestimate of the halo's mass when using spherical Jeans equation modeling.
\end{abstract}


\section{Introduction}
\label{s:intro}

It is widely assumed that the kinematic state of the stellar halo can be used to constrain the Milky Way's (MW) formation history \citep{Eggen1962, Johnston2008} and mass distribution \citep{Xue2008, Gnedin2010, Deason2012}.
As a result, a considerable effort has been expended in measuring the stellar halo's kinematic moments \citep[\textit{e.g.},][]{Xue2008, Bond2010, Cunningham2016}. 
Recently, emphasis has been placed on the measurement of the velocity anisotropy parameter ($\beta$), the ratio of tangential to radial random motion, which is expected to be positive from simple numerical experiments of halo formation \citep{Binney2008}.
However measurements of $\beta$ in the MW have suggested that it is negative within $15 \lesssim R/$kpc $\lesssim 25$ (see \citeauthor{Kafle2012} \citeyear{Kafle2012}, \citeauthor{King2015} \citeyear{King2015}, but see \citeauthor{Deason2013b} \citeyear{Deason2013b}, \citeauthor{Cunningham2016} \citeyear{Cunningham2016} for alternative values), leading to speculation that the exact
merger and dissipation history of the stellar halo could strongly affect its velocity anisotropy profile \citep{Deason2013a, Deason2013b}.
In spite of these recent efforts to infer the MW's accretion history from measurements of the density profile and $\beta$, there have been no systematic studies of how $\beta$ varies with radius in realistic cosmological hydrodynamic simulations that demonstrate that $\beta$ is in fact a tracer of assembly history.

First introduced by \citet{Binney1980} to characterize the orbital structure of a spherical system, $\beta$ is most commonly used in spherical Jeans equation modeling to recover the mass distribution of galactic systems. 
In a Galactocentric spherical coordinate system ($r$, $\theta$, $\phi$), corresponding to radial distance, polar angle, and azimuthal angle, we define $\beta$ as: 
\begin{eqnarray}
\label{eq:eq1}
\beta(r) = 1 - \frac{{\sigma_{\theta}(r)}^2 + {\sigma_{\phi}(r)}^2}{2 \sigma_{r}(r)^2},
\end{eqnarray}
where $\sigma_{\theta}$, $\sigma_{\phi}$, $\sigma_{r}$ are the velocity dispersions in spherical coordinates. 
In a system in which $\beta=1$, all stars are on radial orbits plunging in and out of the galactic center, while in a system with $\beta=-\infty$, all orbits are circular.
A system with an isotropic velocity distribution ($\sigma_{\theta}=\sigma_{\phi}=\sigma_{r}$) has $\beta=0$.

Models of galaxy formation generally imply that $\beta$ increases with radius, corresponding to nearly isotropic near the center and radially biased in the outskirts (see \S4.10.3 of \citeauthor{Binney2008} \citeyear{Binney2008}, and references therein; \citeauthor{Debattista2008} \citeyear{Debattista2008}). 
This trend has been shown in both cosmological pure $N$--body simulations \citep[see Figure 10,][]{Diemand2005} and in cosmological $N$--body+SPH simulations \citep[see Figure 5,][]{Sales2007, Abadi2006}.
Analyzing the $z=0$ snapshot of the high resolution MW-like simulation Eris, \citet{Rashov2013} also found $\beta$ to be increasingly radially biased with distance, transitioning to purely radial stellar orbits beyond $100$ kpc \citep[see \S4 and Figure 2,][]{Rashov2013}.
Notably, Eris shows a ``dip'' in $\beta$ at $r \sim 70$ kpc, where $\beta$ drops from $\sim 0.75$ to $0.5$ over a narrow range of radii, which coincides with  recently accreted substructure \citep[see Figure 3,][]{Rashov2013}.
This hints that fluctuations in the value of $\beta$ are possible in simulations, but does not speak directly to their duration, intensity or frequency of occurrence.
Recently, using orbital integration analysis, \citet{Bird2015} considered the duration of low values in $\beta$ and found them to be short lived (persisting a few tens of Myr) and unconnected to the galactic density profile.
Motivated by this analysis, we look at the time evolution of $\beta$ simulated in a full cosmological context, to understand what, if any, predictive power $\beta$ holds for constraining the formation history of the MW.

From an observational perspective, $\beta$ is hard to measure and somewhat sensitive to small number statistics.
For MW halo stars, the form of the $\beta$-profile measured also depends on the modeling method employed.
Assuming the MW is well described by a truncated, flat rotation curve,
it is possible to derive the velocity anisotropy profile from 4D data (Galactocentric radius, on-sky position and line-of-sight velocity) using an action-based distribution function method \citep[see, for instance,][]{Wilkinson1999, Deason2012}. 
Recently, \citet{Williams2015} constrained such a model using the blue horizontal branch catalog of \citet{Xue2011}. 
Their best fit result for $\beta(r)$ rises appreciably more gradually than $\beta(r)$ from $N$--body simulations \citep[see Figure 8 and \S 5.3][for further details]{Williams2015}.

In contrast, measurements of $\beta$ for halo stars within the solar cylinder based on full 6D phase space data find $\beta$ to be strongly radially biased.
For example, \citet{Chiba1998} analyzed the kinematics of nearby stars falling within $\sim$2 kpc from the Sun.
They leveraged proper motion (and parallax for a handful of stars) from \textit{Hipparcos} satellite and the photometric distance, line-of-sight velocity and [Fe/H] from ground-based telescopes.
Using 124 stars with [Fe/H]$<-1.6$, \citet{Chiba1998} found velocity dispersions $(\sigma_r, \sigma_\phi, \sigma_\theta) \simeq (\sigma_U, \sigma_V, \sigma_W) = (161\pm10, 115\pm7, 108\pm7)$ $\mathrm{km \; s^{-1}}$, corresponding to $\beta=0.52\pm 0.07$. 
Sampling a larger volume (within 5 kpc of the Sun), \citet{Smith2009} found $\beta=0.69\pm0.01$.
This value was determined using a catalog of $\sim$1700 halo subdwarfs selected using a reduced proper-motion diagram applied to SDSS Stripe 82 data; combined with radial velocities from SDSS spectra, and distances from the photometric parallax relation (with uncertainty of $\sim10\%$), \citet{Smith2009} found ($\sigma_{r}$, $\sigma_{\phi}$, $\sigma_{\theta}$) $=$ ($143 \pm 2$, $82 \pm 2$, $77 \pm 2$) km s$^{-1}$. 
Sampling a slightly larger footprint still ($r<$10 kpc) pointed toward the northern Galactic cap, \citet{Bond2010} found  a similar value, $\beta\sim0.67$.
This was determined using proper motions of a large sample of main sequence SDSS stars from \citet{Munn2004} resulting in ($\sigma_{r}$, $\sigma_{\phi}$, $\sigma_{\theta}$) $\sim$ ($141$, $85$, $75$) km s$^{-1}$.

Beyond $r\sim10$ kpc, it has been extremely difficult to obtain full 6D information for a robust sample of halo stars.
Since 2014, the HALO7D project \citep{Cunningham2015} has worked to obtain accurate HST-measured proper motions and very deep Keck DEIMOS spectroscopy of $\sim$100 MW main sequence turn-off stars with the goal of assessing $\beta$ at large radii.
Analysis of 13 HALO7D stars lying within $18<r/$kpc $<32$ yields $\beta=-0.3\substack{+0.4 \\ -0.9}$ \citep{Cunningham2016}.
This value is consistent with isotropy and lower than the solar neighborhood $\beta$ measurements by $2\sigma$.
This value is also substantially lower than model predictions of radially biased values; however, model predictions in the literature were generated in the limit that there was no satellite substructure present.
\citet{Cunningham2016} note that two stars from this sample are likely members of a known substructure (TriAnd).
If they exclude these stars from their analysis, they find $\beta=0.1\substack{+0.4 \\ -1.0}$, which is still formally lower than solar neighborhood measurements but just outside the $1\sigma$ limit.

There is a robust and interesting discussion in the literature of the value of $\beta$ beyond $r\sim20$ kpc based upon 4D phase-space information for thousands of blue horizontal branch stars \citep{Sirko2004, Deason2012, Kafle2012, King2015} and 5D phase-space for a small number of halo stars \citep{Deason2013b}.
Wildly divergent values for $\beta$ have been obtained; based upon these studies, it is plausible that $\beta$ remains radially anisotropic \citep{Deason2012}, $\beta$ ``dips'', falling from a radial $\beta\sim0.5-0.7$ value at r$<$20 kpc to an isotropic $\beta\sim0$ \citep{Sirko2004, Deason2013b} or $\beta$ is strongly tangentially biased, with $\beta<-1.5$ \citep{Kafle2012, King2015} at $r\sim20$--$25$ kpc.
\citet{Deason2013b} speculate that this dip could be associated with a large, shell-type structure that is a remnant of an accretion event at $r\sim25$ kpc; however, \citet{Johnston2008} find shell-type structures to be typically associated with stars on radial orbits at apogalactic passage. 

In a companion paper (Hattori \etal2017), we consider the impact of using 4D data instead of full 6D data to estimate $\beta$.
We find $\beta$ is systematically underestimated beyond a certain radius (r$\sim$15 kpc for the currently available sample size).
As $r$ increases, the line-of-sight velocity approaches the Galactocentric radial velocity.
This makes it difficult to extract information about the tangential velocity distribution (and hence $\beta$) from the line-of-sight velocity distribution alone. 
The limitation of the line-of-sight velocities to recover the velocity anisotropy was first explored in \citet{Hattori2013} and is supported by \citet{Wang2015}, who find that if proper motions are not available, it is difficult to obtain robust constraints on $\beta$.
Thus for the remainder of this paper, we will focus on $\beta$ derived from 6D phase-space information.

We are optimistic that upcoming \textit{Gaia} data will fill in the gaps and tighten constraints on $\beta$(r) for the MW \citep{Gaia2016}.
For example, with a \textit{Gaia} sample of 2000 blue horizontal branch stars within $15<$ r/kpc $<30$ (expected distance error $<5\%$), we anticipate a $\beta$ dip from $0.5$ to $0.0$ is recoverable with an error on beta $<0.2$ (see Appendix for further details).
With this sensitivity in mind, in this paper, we consider what high resolution MW-like simulations predict for $\beta(r)$.
We aim to assemble a comprehensive set of predictions for $\beta(r)$ for observers to reference and challenge in the coming years.
In \S \ref{s:simulation}, we discuss the set-up of the three suites of simulations we use.
In \S \ref{s:discussion}, we present average trends in $\beta(r)$; we find that all three suites are consistent and predict a monotonically increasing value of $\beta$ that is radially biased, and beyond 10 kpc, $\beta > 0.5$.
We also consider $\beta$ as a function of time for individual simulated galaxies, and discuss when and why ``dips'' in $\beta$ form\footnote{Our fiducial definition of a ``dip'' is a value of $\beta$ that is at least $0.2$ lower than $\beta$ at the surrounding radii.} and the rarity of $\beta < 0$ values, the origin and persistence of these dips in the \textit{in situ} and accreted halo.
We also highlight one simulation that is a $\beta(r)$ outlier: while this galaxy appears to be a normal MW-like disk galaxy at present day, it experienced a major merger with a gas rich system at z$\sim$1.
This event left a lasting imprint on the spherically averaged value of $\beta(r)$; the stellar halo has a ``trough'' $\beta$ profile -- a persistently low positive to negative value of $\beta$ over a wide range of radii -- until the present day.
We note this isotropic to tangential $\beta$ feature is not uniform across the sky; however, it is observable in at least a quarter to half of the sky at any given radius.
We speculate that if the MW went through a similar cataclysmic event, then the signature in $\beta$(r) should be visibly present in the MW's stellar halo today and measurable in the foreseeable future.
If, on the other hand, the narrow dip at $r\sim20$ kpc is confirmed or other dips are found, we suggest that these are ideal locations to carry out a follow up search for either substructure or \textit{in situ} halo stars.
These two possibilities can be distinguished by the metallicity and $\alpha$-abundance patterns of the stars giving rise to the $\beta$ dip.
We discuss these results and draw further conclusions in \S \ref{s:conclusion}.

\section{Simulations}
\label{s:simulation}

We analyze 3 different suites of high resolution MW-like stellar halo simulations:  a hybrid $N$--body $+$ semi-analytic suite and two fully $N$--body+SPH suites with differing prescriptions for star formation and stellar feedback.  

\subsection{Bullock \& Johnston Suite}
\label{s:bj05}

We consider 11 stellar halos from \citet{Bullock2005} (henceforth BJ05) which are modeled using the hybrid $N$--body $+$ semi-analytic approach.
These models are publicly available\footnote{found at http://www.astro.columbia.edu/$\sim$kvj/halos/} and are described in detail in \citet{Bullock2005, Robertson2005, Font2006}.

BJ05 assumes a $\Lambda$CDM framework with a $\Omega_{m}=0.3$, $\Omega_{\Lambda}=0.7$, $\Omega_{b}h^2=0.024$, $h=0.7$ cosmology.
They generate 11 merger histories for a $z=0$, $M_{vir}=1.4 \times 10^{12} $ $\mbox{$\rm M_{\odot}$}$ dark matter halo using the method described in \citet{Somerville1999}. 
For each merger event above $5 \times 10^{6} $ $\mbox{$\rm M_{\odot}$}$, an $N$--body simulation of a dark matter satellite disrupting in an analytic, time-dependent galaxy $+$ spherical dark matter halo is modeled.
The baryonic component of each satellite is modeled using semi-analytic prescriptions and the star formation is truncated soon after each satellite halo is accreted on the the MW host.

While BJ05 neglect satellite-satellite interactions and lack a responsive ``live'' halo and central galaxy, their methods have provided robust predictions for the spatial and velocity structure of stellar halos and streams in the outer parts of galaxies ($\ge$ 20 kpc), as well as reasonable estimates for global stellar halo properties from accreted material (mass and time evolution) at all radii \citep{Bell2008}.
Moreover, their models sample a wide range of merger histories within allowable bounds for the MW, which makes them valuable for gaining intuition about the effects of mergers on the phase-space distribution of the stellar halo at present day.

\subsection{g14 Suite}

We use the \textit{g14} \citep{Christensen2012} suite of simulations which contains four cosmologically derived \citep[WMAP3]{Spergel2003} MW–-mass galaxies named \textit{g14\_h239}, \textit{g14\_h258}, \textit{g14\_h277}, and \textit{g14\_h285}; these galaxies are evolved to redshift zero using the parallel $N$--body+SPH code GASOLINE \citep{Wadsley2004}.
These runs have a spatial resolution of 170 pc and mass resolutions of $1.3\times10^5$, $2.7\times10^4$, and $8.0\times10^3$ $\mbox{$\rm M_{\odot}$}$ for the dark matter, gas and stars, respectively, while also including the large-scale environment by using the `zoom-in' volume renormalization technique \citep{Katz1993} to create the initial conditions. 
The simulations use a redshift dependent cosmic UV background and realistic cooling and heating, including cooling from metal lines \citep{Shen2010}.
Supernovae feedback is modeled using the ``blastwave'' approach \citep{Stinson2006} in which cooling is temporarily disabled based on the local gas characteristics.
The probability of star formation is a function of the non-equilibrium $H_2$ abundances \citep{Christensen2012}.
The result of tying the star formation to the molecular hydrogen abundance is a greater concentration of the stellar feedback energy and the more efficient generation of outflows.
These outflows ensure that the final galaxies have appropriate rotation curves \citep{Governato2012}, stellar mass fractions \citep{Munshi2013}, and dwarf satellite populations \citep{Zolotov2012, Brooks2014}.

\begin{figure*}[!ht]
  \hskip -.3in
  \epsscale{1}
  \plotone{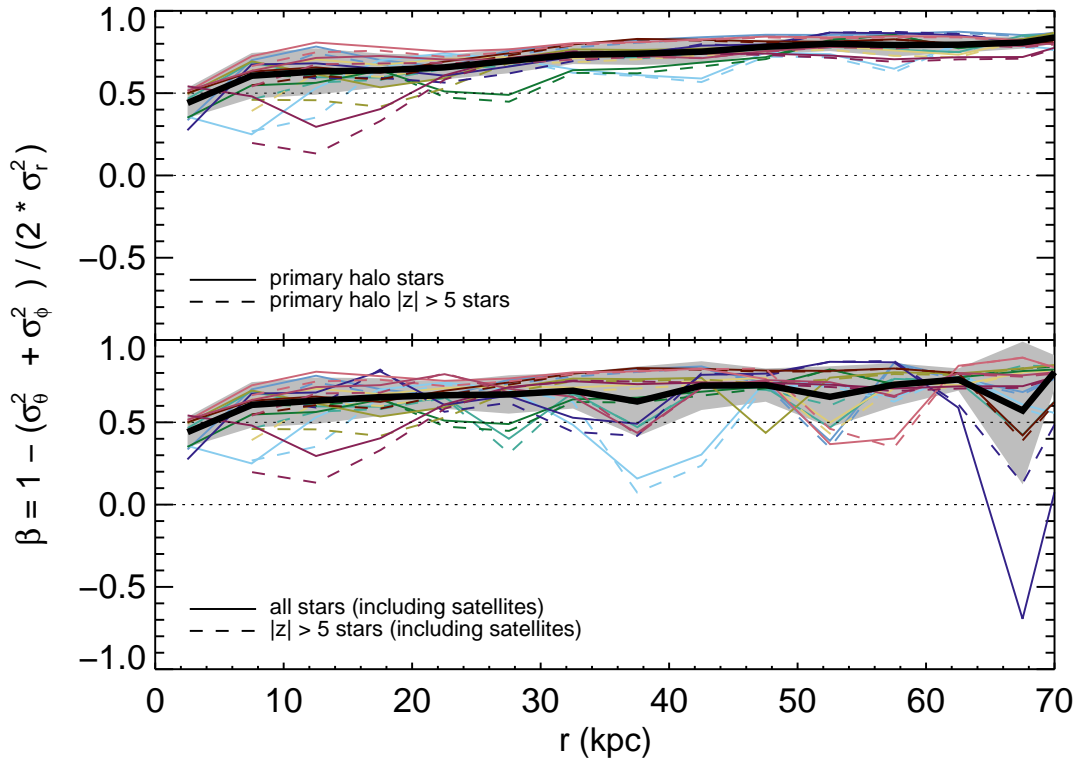}
  \caption{$\beta$ as a function of radius for 11 stellar halos from BJ05. Top panel: $\beta$ for stars belonging to primary halo at present day.  Thin lines correspond to individual halos, thick line corresponds to average behavior, shaded gray shows area within 1$\sigma$ of the mean.  Solid lines correspond to all stars and dashed lines correspond to $|z| > 5$ kpc stars. Bottom panel:  $\beta$ for all stars within the virial radius (including satellites) at present day.  The thin, thick, solid, and dashed lines and shaded region are the same as in the top panel.}
  \label{f:bj}
\end{figure*}

\subsection{\textit{MaGICC} Suite}

We utilize 2 cosmological hydrodynamic simulations named \textit{MaGICC\_g1536} and \textit{MaGICC\_g15784}, from the Making Galaxies in a Cosmological Context \citep[\textit{MaGICC},][]{Stinson2012} suite of simulations.
Like the $g14$ suite, the \textit{MaGICC} galaxies were generated using GASOLINE \citep{Wadsley2004}; however, instead of disabling cooling at early times, the \textit{MaGICC} implementation includes early stellar feedback from massive stars, which is purely thermal and operates much like an ultraviolet ionization source.
The early heating of the gas suppresses a higher fraction of star formation prior to $z=1$ than supernovae feedback alone; thus, the \textit{MaGICC} galaxies do not suffer from overcooling, and have realistic rotation curves \citep[see Figure 1,][]{Santos-Santos2016} with smaller central bulges in the MW host galaxies and more realistic stellar content in satellite galaxies.

The \textit{MaGICC} simulations contain dark matter, gas and star particles with masses of $1.11 \times 10^6$ $\mbox{$\rm M_{\odot}$}$, $2.2 \times 10^5$ $\mbox{$\rm M_{\odot}$}$, and $< 6.3 \times 10^4$ $\mbox{$\rm M_{\odot}$}$, respectively, and a gravitational softening length of 310 pc. The two \textit{MaGICC} galaxies we analyze, \textit{MaGICC\_g1536} and \textit{MaGICC\_g15784} have been studied extensively previously \citep[see][and references therein]{Snaith2016}.

For both the \textit{MaGICC} and \textit{g14} simulations, halo membership is determined using the density-based halo finding algorithm AHF \citep{AHF2004, AHF2009}.
We previously analyzed the \textit{in situ} and accreted stellar halo from \textit{MaGICC\_g15784} in \citet{Valluri2016}; in this work, any star belonging to the primary halo at present day is classified either as an \textit{in situ} star or accreted star.
Stars that are born in the primary halo are classified as \textit{in situ} stars, while stars that are born in other bound structures are classified as accreted.
Because we are interested in the kinematic properties of the stellar halo, we distinguish between \textit{in situ} halo and \textit{in situ} disk stars based purely on a spatial cut; at present day, any \textit{in situ} stars with $|z|>5$ kpc are considered \textit{in situ} halo stars.\footnote{Throughout this work, we orient each simulated MW-like galaxy with its angular momentum vector pointed along the $z$-axis, ensuring its disk is aligned in the $x$-$y$ plane.}
For \textit{MaGICC\_g1536}, 24\% of the halo stars are \textit{in situ} halo stars, and for the more massive system, \textit{MaGICC\_g15784}, 42\% of the halo stars are \textit{in situ} halo stars.

\section{Results}
\label{s:discussion}

\subsection{Radially Anisotropic Trends}

\begin{figure*}[!ht]
  \hskip -.3in
  \epsscale{1}
  \plotone{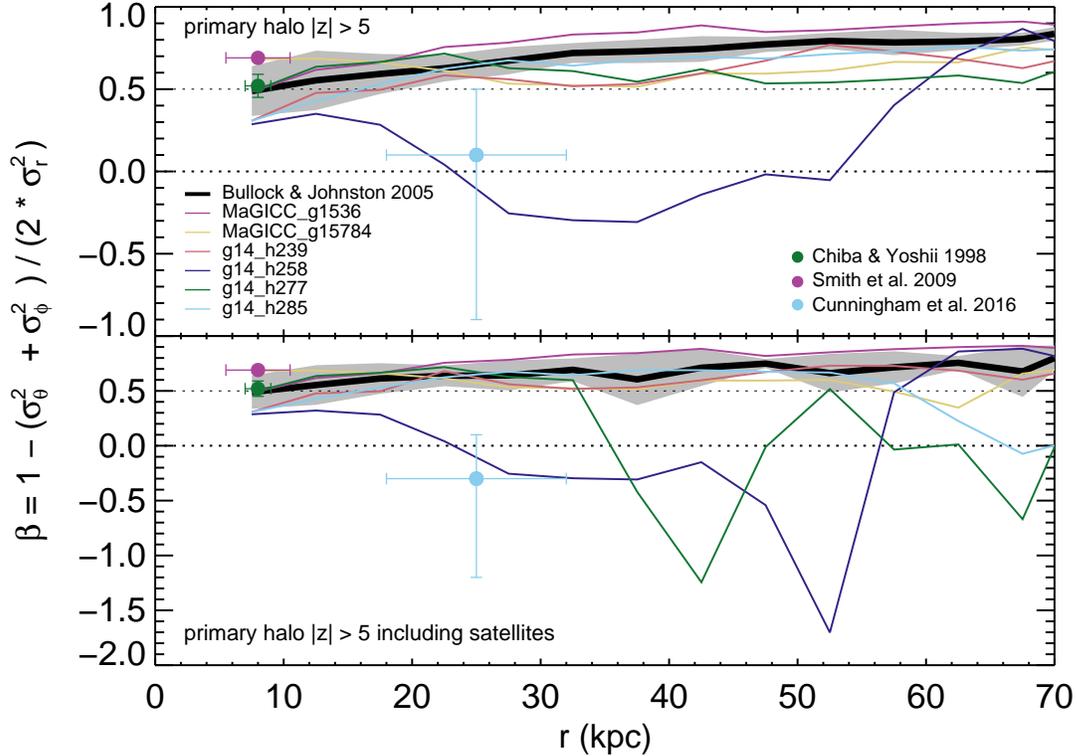}
  \caption{
    Top panel: $\beta(r)$ profiles for all stars belonging to the primary stellar halos from six cosmological simulations (colored lines) and the average profile from BJ05 simulations (thick black curve), with the region within 1$\sigma$ of the mean for BJ05 shown in gray.
    Only one galaxy ($g14\_h258$) shows significant (negative) deviation from the average curve over a large range of radii (see \S\ref{s:h258} for details). The three points mark three measurements of $\beta$ in the MW from 6D coordinates. Bottom panel: same as above for all stars including those bound to satellites within the virial radius at present day.}
  \label{f:z0}
\end{figure*}

We begin by considering the $z=0$ behavior of $\beta(r)$ for the BJ05 suite of simulations.
As noted in \S\ref{s:bj05}, the BJ05 models are produced using a hybrid $N$--body $+$ semi-analytic approach which results in stellar halos formed purely from accreted material.
Henceforth, we adopt a fiducial radial bin size of 5 kpc.

The top panel of Figure~\ref{f:bj} presents $\beta(r)$ for the BJ05 models for stars that belong to the primary halo at present day.
The behavior of each individual halo is shown by thin lines, while the average behavior of all 11 halos is shown in the thick black solid line surrounded by the $1\sigma$ error bands shaded in gray.
As a check, we look at $\beta(r)$ for two cuts on the data: all the stars in the stellar halo (solid line), and just the stars with $|z|>5$ kpc (dashed line).
There is no difference in the average $\beta(r)$ values for these two populations.
As also shown in \citet{Williams2015}, in the BJ05 suite, the average trend is quite radially biased at all radii.
From the smallest radial bin outward, $\beta\ge0.5$; by $r\sim30$ kpc, $\beta\sim0.7$, and for larger $r$, $\beta$ asymptotes to $\sim0.8$.
Regardless of merger history, all 11 halos show the same global behavior, trending toward large values of $\beta$ at large $r$.
In fact, the halo with a large late time accretion event (halo 9, shown in green) is relatively indistinguishable in $\beta$ from the other 10 halos. 
While there are slight dips in $\beta$ for individual halos, these dips never plummet to tangential or even isotropic values.
Most dips are fairly small (on order $0.2$ to $0.3$ lower than the average $\beta$ value), and beyond $r\sim20$ kpc, even these dips do not descend below $\beta\sim0.5$.

The bottom panel of Figure~\ref{f:bj} presents $\beta(r)$ for all stars in the simulation, including those bound to infalling satellites.
Because stars in a satellite lie within a small spatial volume and follow a coherent trajectory, including satellites generates dips in individual $\beta(r)$ profiles; these dips are $\sim5-15$ kpc wide.
A significant number of these dips fall below $\beta\sim0.5$; however, unexpectedly, very few of the dips could be considered isotropic and only one is tangential.
Moreover, in the tangential instance (dark blue curve), it is very clear that the stars generating the dip belong to a small, coherent structure; this can be seen by the substantial difference in $\beta$ for the full sample and the $|z|>5$ kpc sample at $r\sim65$ kpc.

We consider next the individual trends in the six $N$--body+SPH simulations from the \textit{g14} and \textit{MaGICC} suites.
Here we select stars belonging to the stellar halo by a spatial cut ($|z| > 5$ kpc).
The top panel of Figure~\ref{f:z0} shows $\beta(r)$ for the stars that belong to the primary halo at the present day.
Plotted in black is the average trend from Figure~\ref{f:bj}, with the 1$\sigma$ error band plotted in gray.
Five of the six galaxies follow the BJ05 trend: from r$\sim10-15$ kpc onward, $\beta \ge 0.5$.
For these galaxies, $\beta$ never falls below $0.5$ and generally trends toward larger values of $\beta$ with increasing radius.
While these five galaxies represent a wide range of merger histories for $z<1$, their $\beta(r)$ behavior is remarkably consistent with one another:
\textit{g14\_h239} (shown in salmon) has the most active merger history and yet its $\beta(r)$ is virtually indistinguishable from \textit{g14\_h277} (shown in green) which has a remarkably quiescent merger history until the very end of the simulation.
Interestingly, the one galaxy that does not follow the BJ05 trend, \textit{g14\_h258}, has a somewhat unremarkable merger history for $z<1$.
We will discuss this galaxy further in \S$3.2.3$. 
It is remarkable, though, that none of the simulations' minor mergers from $z<1$ leave a lasting impression on $\beta(r)$.
$\beta$ is predicted by the average trends in BJ05, \textit{g14}, and \textit{MaGICC} suites to be $\sim0.5$ or larger at all radii beyond $r\sim8$ kpc at present day.

The three individual data points on Figure~\ref{f:z0} mark existing measurements based on 6D data in the MW from nearby stars falling within $\sim$2 kpc from the Sun \citep{Chiba1998}, SDSS stars in Stripe 82 with 5 kpc of the Sun \citep{Smith2009}, and from 13 HALO7D stars lying within $18<r/$kpc $<32$ \citep{Cunningham2016}.
Note that the measurements of anisotropy from nearby stars (forest green point) and SDSS (salmon point) are completely consistent with predictions from all the simulations.
The error bars on the measurement from HALO7D (pale blue point) are large but the measured value, while still positive in the top panel, is significantly lower than the predictions from most of the simulations and intriguingly is consistent with the predictions from \textit{g14\_h258}.

The bottom panel of Figure~\ref{f:z0} presents $\beta(r)$ for all stars in the simulation inside the virial radius of the primary halo but with $|z| > 5$ kpc, including those bound to infalling satellites.
Here, it is obvious that three of the six galaxies are interacting with satellites at present day: \textit{g14\_h277}, \textit{g14\_h285} and \textit{MaGICC\_g15784}.
The first two of these galaxies have strongly tangential dips in $\beta$.
These dips are much stronger than the tangential dip seen in BJ05.
The dip in \textit{MaGICC\_g15784} ($\beta\sim0.4$) is still a radial value, but it would be stronger if the satellite was aligned differently with the disk, as it falls within $|z| < 5$ kpc at the end of the simulation.

\begin{figure*}[!ht]
  \hskip -.3in
  \epsscale{1}
  \plotone{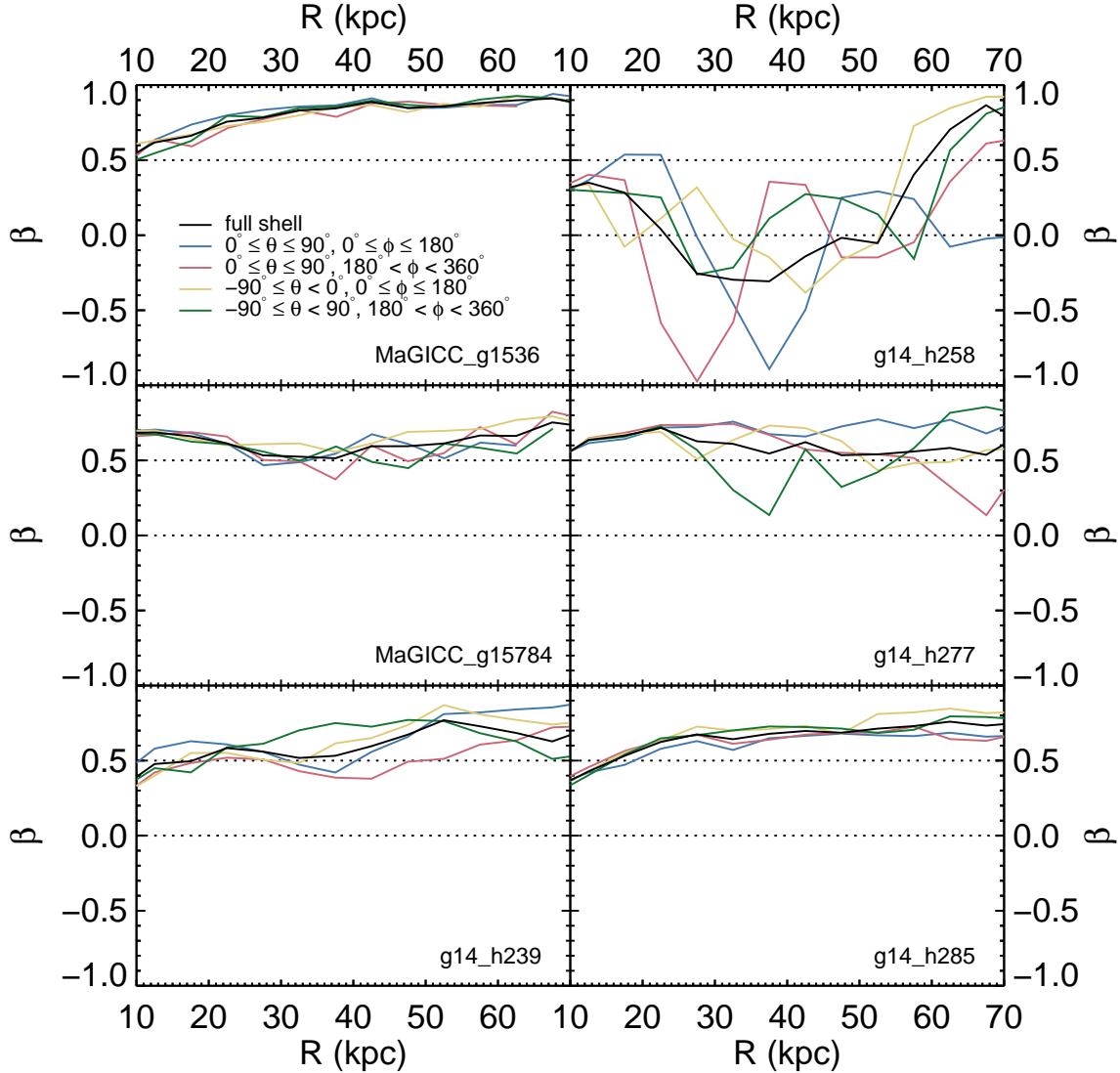}
  \caption{The $\beta(r)$ profiles by angular quadrants for the six cosmological hydrodynamic simulations considered in Figure~\ref{f:z0}.  For each galaxy, the total $\beta(r)$ profile is shown in black, and the $\beta(r)$ profiles for the ($0^{\circ} < \theta < 90^{\circ}, 0^{\circ} < \phi < 180^{\circ}$), ($0^{\circ} < \theta < 90^{\circ}, 180^{\circ} < \phi < 360^{\circ}$), ($-90^{\circ} < \theta < 0^{\circ}, 0^{\circ} < \phi < 180^{\circ}$), and ($-90^{\circ} < \theta < 0^{\circ}, 180^{\circ} < \phi < 360^{\circ}$) angular quadrants are shown in blue, salmon, yellow and forest green respectively.  With the exception of \textit{g14\_258}, the total behavior closely mimics the quadrant behavior except where a galaxy is actively accreting a satellite (as in the case of \textit{g14\_h277}).  Outlier \textit{g14\_258} shows a complex angularly and radially dependent $\beta(r)$ signature, which we discuss in further detail in \S\ref{s:h258}.}
  \label{f:ang}
\end{figure*}

Building on this, we next explore how uniform $\beta$ is across the sky.
Figure~\ref{f:ang} shows the $\beta(r)$ profiles for the six hydrodynamic simulations from Figure~\ref{f:z0}, but now subdivided by angular quadrants.
The four non-overlapping angular quadrants that we consider are ($0^{\circ} < \theta < 90^{\circ}, 0^{\circ} < \phi < 180^{\circ}$), ($0^{\circ} < \theta < 90^{\circ}, 180^{\circ} < \phi < 360^{\circ}$), ($-90^{\circ} < \theta < 0^{\circ}, 0^{\circ} < \phi < 180^{\circ}$), and ($-90^{\circ} < \theta < 0^{\circ}, 180^{\circ} < \phi < 360^{\circ}$).
For each galaxy, the total $\beta(r)$ profile from Figure~\ref{f:z0} is shown in black, while the $\beta(r)$ profile for each angular quadrant is shown in a colored line.

At a given radius is $\beta$ the same in every direction we look?
By and large, yes, it is the same in every direction we look for the five ``typical'' hydrodynamic simulations.
The total behavior closely mimics the quadrant behavior except where a galaxy is actively accreting a satellite, as in the case of \textit{g14\_h277}.
However, the one outlier galaxy, \textit{g14\_258}, shows a complex angularly and radially dependent $\beta(r)$ signature.
We discuss this galaxy in further detail in \S\ref{s:h258}.
Overall, we conclude that unless a galaxy is actively accreting a satellite or experienced a unique cataclysmic merger, at a given radius, $\beta$ is self consistent across the sky.

We conclude from this analysis of 17 $z=0$ MW-like stellar halos that, except i the rarest of cases, $\beta(r)$ is strongly predicted to be radially anisotropic beyond $r\sim8$ kpc.
In fact, the average trends for all three suites of simulations predict that $\beta\sim0.5$ or larger at all radii at present day.

\subsection{Deviations from Radial Anisotropy}

\begin{figure*}[!htb]
  \hskip -.3in
  \epsscale{1}
  \plotone{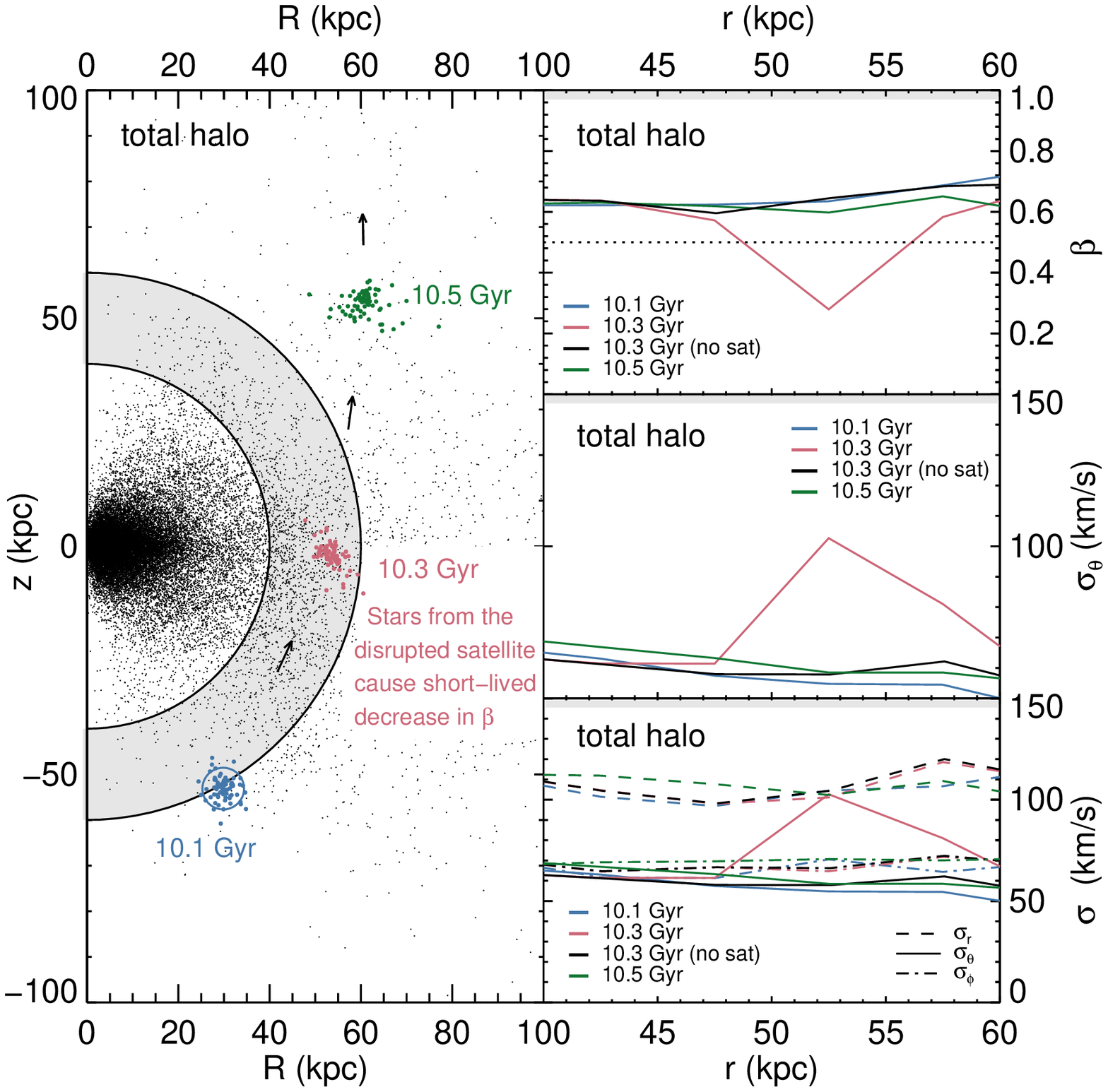}
  \caption{Illustration of the formation of a short-lived ``dip'' in $\beta$ in the total (accreted$+$\textit{in situ}) stellar halo of \textit{MaGICC\_g15784}. Left: The black points show the total stellar halo (in cylindrical coordinates) at $10.5$ Gyr. The gray shaded region between the solid curves marks the radial shell for which $\beta$ and $\sigma$ profiles are shown in the other panels. Colored points mark the location of stars belonging to a disrupting satellite as it passes through the MW-like galaxy \textit{MaGICC\_g15784} at three different times (blue, red, dark green corresponding to time $\sim$10.1, 10.3, and 10.5 Gyr); the circle on top of the stars at 10.1 Gyr indicates that the satellite is bound at this time.  All of these stars are identified as belonging to \textit{MaGICC\_g15784}'s stellar halo by time $\sim$10.3 Gyr. The black arrows mark the trajectory of the stars as the satellite breaks up. Top right: $\beta(r)$ profiles for total stellar halo at three different times (with and without the stars from the disrupted satellite at $t=10.3$~Gyr). Middle right: the corresponding polar velocity dispersion. Bottom right: all three components of the velocity dispersion for the total halo stars.}
  \label{f:total}
\end{figure*}

As we have shown, a robust prediction of $\Lambda$CDM simulations is that stellar halos are radially anisotropic ($\beta \ge 0.5$).
However, recent analysis of 6D MW data indicates a low value of $\beta$ at larger radii in our galaxy \citep{Cunningham2016}; these observations prompt us to explore when and how rare departures from radial anisotropy occur in simulations.
In what follows, we conduct a time series investigation of two hydrodynamic simulations, \textit{MaGICC\_g15784} and \textit{g14\_h258}; we explore three different scenarios when deviations from radial anisotropy occur:
\begin{enumerate}
\item{An ongoing accretion event can cause a short-lived ($\Delta\textrm{time}<0.2$ Gyr) dip in $\beta$ over a small range in radii.}
\item{Close passage of a large satellite galaxy can cause a longer-lived ($\Delta\textrm{time} \ge 0.4$ Gyr) $\beta$ dip in the \textit{in situ} halo over a small range in radii.}
\item{A major merger event can cause a very long-lived ($\Delta\textrm{time} \sim 7$ Gyr) tangential $\beta$ feature across a large range of radii and a large angular fraction of the sky.}
\end{enumerate}
Here we illustrate each of these scenarios in turn.

\subsubsection{Transient $\beta$ Dips in the Total Stellar Halo}

We now consider the total stellar halo for \textit{MaGICC\_g15784}, which is dominated by accreted stars beyond $r\sim30$ kpc and is slightly oblate with a short/long axis ratio $c/a \sim 0.85$.
At $z=0$, \textit{MaGICC\_g15784} has a virial radius, virial mass, and stellar mass of $R_{200} = 214$ kpc, $M_{200} = 1.2\times10^{12}$ $\mbox{$\rm M_{\odot}$}$, and $M_{*} = 8.3\times10^{10}$ $\mbox{$\rm M_{\odot}$}$ respectively,\footnote{Here we have defined the virial radius to be $R_{200}$, the radius at which the average density of the halo is 200 times the critical density of the Universe, and the virial mass to be the total mass within the virial radius.} and experienced its last major merger at z $\sim 1$. 

The left panel of Figure~\ref{f:total} illustrates the spatial distribution (in Galactocentric cylindrical coordinates, $z$ versus $R$) of stars in the stellar halo at time $\sim 10.5$ Gyr (redshift $z \sim 0.3$).
The gray shaded region corresponds to a spherical shell spanning $40<r/$kpc$<60$ containing a stellar halo mass of $7.2\times10^{7}$ $\mbox{$\rm M_{\odot}$}$, which we look at in detail in the other three panels of Figure~\ref{f:total}.
At $10.1$ Gyr, a bound satellite (stars shown in blue) enters the gray shaded region; this satellite contains a total stellar mass of $2.6\times10^6$ $\mbox{$\rm M_{\odot}$}$.
The black arrows show the direction of movement of the satellite.
As it moves up through the mid-plane, the satellite is disrupted and no longer identified by the halo finding algorithm as a unique object.  
However, stars from this satellite maintain coherence for several time-steps, as illustrated by the location of these stars at $10.3$ and $10.5$ Gyr (shown in red and dark green in the top left panel of Figure~\ref{f:total}). 

\begin{figure*}[!ht]
  \hskip -.3in
  \epsscale{1}
  \plotone{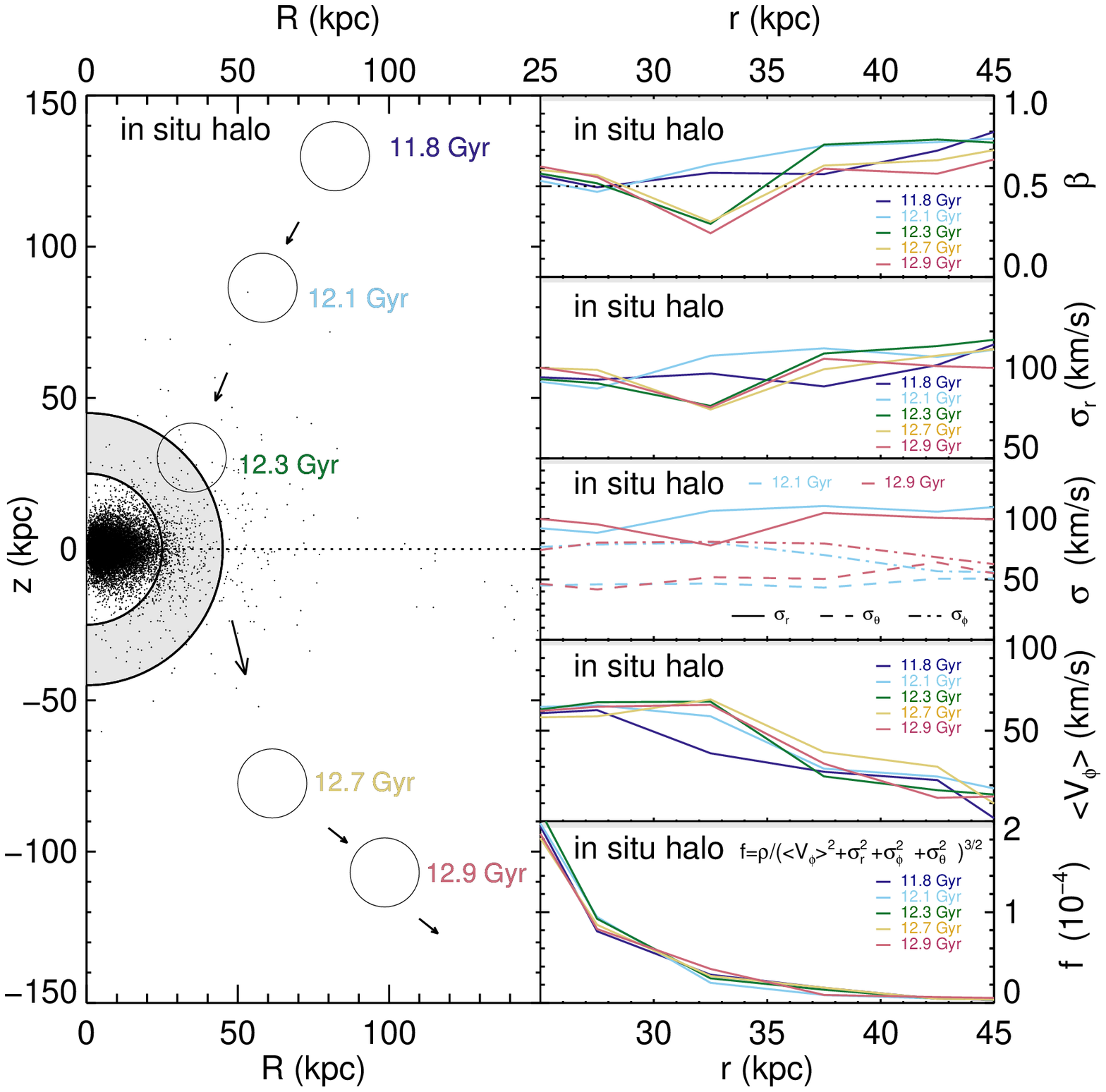}
  \caption{Illustration of the formation of a long-lived ``dip'' in $\beta$ in the \textit{in situ} stellar halo of \textit{MaGICC\_g15784}.  Left: The black points show the \textit{in situ} stellar halo (in cylindrical coordinates) at $12.3$~Gyr. The gray shaded region between solid curves marks the radial shell for which $\beta$ and $\sigma$ profiles are shown in other panels. The open circles correspond to the position of a large $M_{total}=4.4 \times 10^{10} $ $\mbox{$\rm M_{\odot}$}$ satellite (roughly twice the mass of the Small Magellanic Cloud \citep{Besla2012}) at five different times as indicated by the labels; black arrows mark the trajectory of the satellite. Top right: $\beta(r)$ profiles for the  \textit{in situ} stellar halo at the five different times. Second from the top right: the corresponding radial velocity dispersion.  Middle right: all three components of velocity dispersion for the \textit{in situ} halo stars before and after the satellite interaction. Second from the bottom right: the corresponding mean azimuthal velocity profile at all five times. Bottom right: the (pseudo) coarse-grained phase-space density quantity, $f=\rho/(\langle{v_\phi}\rangle^2+\sigma_{r}^2+\sigma_{\phi}^2+\sigma_{\theta}^2)^{3/2}$ in units of $10^{-4}$.  Note $f$ is $\sim$ constant with time, suggesting that this proxy for phase-space density is conserved.}
  \label{f:insitu}
\end{figure*}

The top right panel of Figure~\ref{f:total} shows $\beta(r)$ for $10.1, 10.3$ and $10.5$ Gyr for stars falling within $40<r/$kpc $<60$ and belonging to the total stellar halo at those time-steps.
At $10.1$ Gyr, the bound satellite enters the $40<r/$kpc $<60$ shell.
The $\beta$ anisotropy at $10.1$ Gyr is greater than $0.6$ at all radii within the volume; at this time, the stars that belong to the satellite are not considered a part of stellar halo, and thus $\beta(r)$ is not impacted by it.
However, by $10.3$ Gyr, the satellite has fully disrupted and stars from it are now considered a part of the total stellar halo; at this time a strong dip to $\beta\sim0.25$ appears at $50<r/$kpc $<55$ (shown in red).
This dip arises because stars from the disrupted satellite, which now lie inside this radial range, are on a polar orbit (as seen in the left panel) and hence their net orbital motion adds to the dispersion in the $\theta$ direction.
The former satellite's contribution can be seen clearly by contrasting $\beta(r)$ for all the stars in the stellar halo (red line) to $\beta(r)$ excluding the former satellite's stars (black line).
The black line is greater than $\sim0.6$ at all radii, just like $\beta(r)$ at $10.1$ Gyr.
At $10.5$ Gyr, $\beta(r)$ in no longer impacted by the former satellite in the range of radii under consideration, as the stars from the former satellite have moved out of the spherical shell. 

Why does a dip form with the addition of the recently stripped stars?
The middle right panel of Figure~\ref{f:total} shows $\sigma_{\theta}$ at $10.1$ (blue line), $10.3$ (salmon line), and $10.5$ (forest green line) Gyr.
Clearly,  $\sigma_{\theta}$  is substantially enhanced by adding the satellite stars; however, $\sigma_{\phi}$, $\sigma_{r}$ remain unchanged (see bottom right panel of Figure~\ref{f:total}).
This is because (as can be seen in the left panel) the satellite is on a predominantly polar orbit and hence the satellite debris has a large $v_{\theta}$.
Again, when we remove stars from the disrupted satellite at $10.3$ Gyr (black line) the dip in $\sigma_\theta$ disappears, confirming that this coherent substructure is the source of the dip in $\beta$.
Note, in this instance, an inspection of the stellar halo's $v_{\theta}$ distribution indicates the presence of the satellite debris with a slight overdensity of stars at larger values of $v_{\theta}$.
However, even in this case, we emphasize that $\beta(r)$ is an instructive complementary tool, which allowed us to quickly hone in on an interesting radial bin with minimal effort.

We track the disrupted satellite for several more time-steps and find the $\beta$ dip does occur at larger radii, albeit to a lesser extent.
This is because, as the disrupted satellite continues on its original orbit, it becomes increasingly radial.
We note that this does not explain why the recently accreted stars do eventually turn radially anisotropic; however, the particulars of that transition are outside of the scope of this paper to explore.

We conclude that dips in $\beta$ generated in the total stellar halo are short-lived (lifetime $< 0.2$ Gyr) and closely tied to recent accretion events.
We suggest that hunting for such dips in velocity anisotropy, particularly at large radii, may be an effective means for identifying recently accreted but somewhat dispersed material.

\subsubsection{$\beta$ Dips in the \textit{In Situ} Stellar Halo}

We now consider \textit{MaGICC\_g15784}'s \textit{in situ} stellar halo within $25<r/$ kpc $<45$ between $11.8$ and $12.9$ Gyr.
As noted in \S$2.3$, \textit{in situ} halo stars are distinguished from \textit{in situ} disk stars by a spatial cut at $z=0$.
At $11.8$ Gyr, the number of \textit{in situ} and accreted halo stars are roughly equal at $25$ kpc, although the \textit{in situ} stars are more concentrated toward the plane of the disk.
Their kinematic behavior is also different; we see evidence of this in the response of the \textit{in situ} stellar halo to the passage of a large, gas-rich satellite ($M_{total}=4.4 \times 10^{10} $ $\mbox{$\rm M_{\odot}$}$, roughly twice the mass of the Small Magellanic Cloud \citep{Besla2012}) through the volume at $12.3$ Gyr.

The left panel of Figure~\ref{f:insitu} illustrates the spatial distribution (in Galactocentric cylindrical coordinates, $z$ versus $R$) of stars in the \textit{in situ} stellar halo at time $\sim 12.3$ Gyr.
The gray shaded region corresponds to a spherical shell spanning $25<r/$kpc$<45$ which we look at in detail in the other five panels of Figure~\ref{f:insitu}.
The unfilled circles show the location of the large satellite that passes through the volume at $11.8$, $12.1$, $12.3$, $12.7$ and $12.9$ Gyr with black arrows indicating the direction of motion over time.
At $12.3$ Gyr, the satellite begins its passage through the region in question, but by $12.7$ Gyr, the satellite has moved beyond the relevant volume.
Note, no stars are donated by the satellite to the stellar halo during this passage, nor would such an exchange impact the \textit{in situ} stellar halo, as \textit{in situ} stars are by definition produced only in the primary halo.

\begin{figure*}[!ht]
  \hskip -.3in
  \epsscale{1}
  \plotone{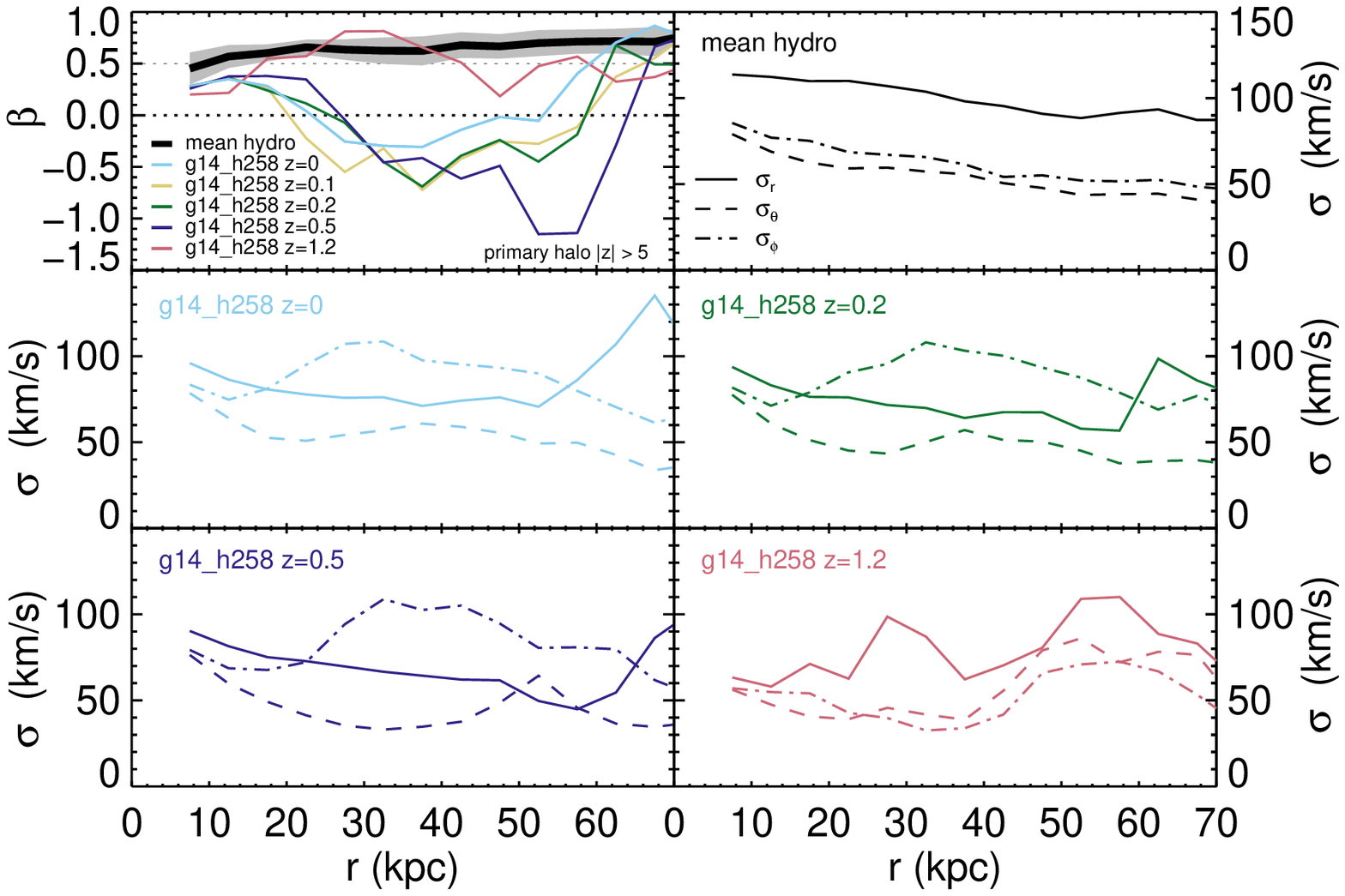}
  \caption{Signatures of major merger in \textit{g14\_h258}. Top left: $\beta(r)$ profiles at five different time-steps compared with the mean $\beta(r)$ for the halo stars from the other five hydrodynamic simulations. The remaining panels show radial profiles of $\sigma_r$, $\sigma_\theta$ and $\sigma_\phi$ for the mean of the 5 hydrodynamic simulations (top right) and for four other the time-steps. The middle panels and bottom left panel correspond to times after the merger event and the bottom right panel corresponds to a time before the merger event.}
  \label{f:h258}
\end{figure*}

The top right panel of Figure~\ref{f:insitu} shows $\beta$ for all five moments in time for the \textit{in situ} stars from the gray shaded region.
Note, we require at least 20 star particles within each radial bin to calculate $\beta$ and within $30<r/$kpc $<35$ there are at least 125 \textit{in situ} halo star particles at each time-step.
Before the satellite interacts with \textit{MaGICC\_g15784}, $\beta$ for the \textit{in situ} stellar halo is consistent with the average behavior of BJ05; as can be seen by the dark and light blue lines for $\beta$ at 11.8 and 12.1 Gyr respectively, $\beta$ is either $\sim0.5$ or larger at all radii in question and is as high at $\sim0.7$ between $30<r$/kpc $<35$ at 12.1 Gyr.
However, starting at $12.3$ Gyr onward, $\beta$ sharply dips to $0.2-0.3$ between $30<r$/kpc $<35$.
This dip persists until the present day.
Other such long lasting \textit{in situ} $\beta$ dips are found elsewhere in \textit{MaGICC\_g15784} and \textit{MaGICC\_g1536} and are coincident with the passage of a $\sim 1.0 \times 10^{9}$ $\mbox{$\rm M_{\odot}$}$ satellite through the $z=0$ plane; however, in all these other cases, the \textit{in situ} $\beta$ dips are radially anisotropic ($\beta > 0$).

The kinematically hotter accreted stellar halo does not experience a similar dip in $\beta$ at this radius at this epoch.
Then why does the \textit{in situ} $\beta$ dip form and persist in this case?
As can be seen in the second from the top panel on the right of Figure~\ref{f:insitu}, $\beta$ declined within $30<r$/kpc $<35$ because $\sigma_{r}$ decreases at $12.3$ Gyr.
However, as can be seen in the middle right panel of Figure~\ref{f:insitu}, neither $\sigma_{\phi}$ nor $\sigma_{\theta}$ are appreciably altered.
At the same time, it is clear that there is an increase in the mean streaming motion in this volume $\langle{v_\phi}\rangle$ (see the second from the bottom right panel of Figure~\ref{f:insitu}).
This increase appears to result from torquing on the \textit{in situ} halo stars originating from the passage of the massive satellite which imparts angular momentum to the \textit{in situ} halo stars.
During the encounter the satellite (which is moving retrograde relative to \textit{g15784}'s disk's rotation) loses orbital angular momentum.
The increase in angular momentum of the \textit{in situ} halo stars results in a corresponding decrease in $\sigma_{r}$.
Note, we have computed the pseudo phase-space density, $\rho/(\langle{v_\phi}\rangle^2+\sigma_{r}^2+\sigma_{\phi}^2+\sigma_{\theta}^2)^{3/2}$, for the \textit{in situ} halo stars (bottom right panel of Figure~\ref{f:insitu}), and it is clear that the radial profile of this quantity does not change during the interaction.
This constancy in the coarse grained phase space density profile is reminiscent of the Liouville theorem although we caution that $f$ is not the fine-grained phase space density, to which the Liouville theorem applies.
This suggests and the reason that the dip in $\beta$ persists in this case is that the  stars contributing to the dip have had their kinematic  and density distributions permanently altered in a way that results in a long term equilibrium. 

As noted in earlier studies, the $\textit{in situ}$ stellar halo is on average more metal-rich and has a lower $\alpha$-abundance than the accreted stellar halo \citep{Zolotov2009,Tissera2013,Pillepich15}.
In \citet{Valluri2016} and Loebman \etal(in prep), we analyze the ages, metallicity, and orbits of accreted and \textit{in situ} stellar halos in the \textit{MaGICC} suite, and we also find that the $\textit{in situ}$ halo stars are on average more metal-rich (on average 0.7 dex higher metallicity) and have a lower $\alpha$-abundance than the accreted halo stars in the same volume.
While our detailed analysis of the connection between metallicity and \textit{in situ} origin is forthcoming, we speculate that if $\beta$ dips are identified in observational data-sets, then metallicity could be used to help distinguish their origin. Did these halo stars form in a small satellite that was recently disrupted?  This accretion origin would correspond to a low to average metallicity in the stellar halo at this radius.  Or did they form in the Milky Way?  This \textit{in situ} origin would corresponds to a higher metallicity than stars at neighboring radii in the stellar halo.

\subsubsection{Merger Induced $\beta$ Trough}
\label{s:h258}

We consider now the $\beta(r)$ outlier, \textit{g14\_h258}, shown in dark blue in Figure~\ref{f:z0} and the top right panel in Figure~\ref{f:ang}.
Like the other galaxies in the  \textit{g14} suite, \textit{g14\_h258} is a good proxy for the MW at $z=0$ by total mass, total stellar mass, and bulge-to-disk ratio \citep{Governato2009, Christensen2012}; however, as discussed in \citet{Governato2009}, \textit{g14\_h258} experiences a major merger (mass ratio of merging halos $1.2:1$) at $z\sim1$.
At this time, the progenitor galaxies plunge in on fairly radial orbits, with the internal spins of the two disks roughly aligned with the orbital angular momentum vector \citep[see Figure~1a][]{Governato2009}.
Over 1 Gyr, the progenitors experience two close passages, and finally coalesce at $z\sim0.8$, thickening the stellar disks and populating the stellar halo in the process.
From $z\sim0.8$ onward, the system has a relatively quiescent merger history as it regrows its thin disk through accreted gas.

In the top left panel of Figure~\ref{f:h258}, we consider $\beta(r)$ over time for \textit{g14\_h258}.
Before the major merger occurs at $z\sim1.2$ (shown in salmon), $\beta(r)$ is consistent with the average profile for BJ05 for $r < 30$ kpc.
While $\beta(r)$ does show a dip at $r \sim 45$ kpc due to a satellite interaction, this dip is minor (neither isotropic nor tangential).

However, for every time-step for $z < 1$, $\beta(r)$ shows a tangential to isotropic profile over a wide range of radii.
That is, the imprint of $z \sim 1$ merger event is encoded in the orbits of the halo stars.
This can be seen clearly in the trends for each component of the velocity dispersion as a function of radius.
The average trends for the five ``normal'' $N$--body+SPH galaxies from Figure~\ref{f:z0} are shown in the top right hand panel of Figure~\ref{f:h258}; here at all radii $\sigma_r > \sigma_{\phi} > \sigma_{\theta}$.
However, in the middle two panels and bottom left panel of Figure~\ref{f:h258}, $\sigma_{\phi} > \sigma_{r} > \sigma_{\theta}$ over the radii in which $\beta(r) \leq 0$.
This is due to a significant enhancement in $\sigma_{\phi}$ and a minor cooling/suppression in growth of $\sigma_{r}$.

Physically, why does this happen?
A detailed analysis of velocity dispersion profiles and $\beta$-profiles in four different quadrants of the galaxy \textit{g14\_h258} at $z=0$ reveals that the trough in $\beta$ results from multiple substantially narrower dips in $\beta$ each only about 10-30~kpc wide.
Furthermore, each quadrant exhibits two distinct dips (see the top right panel of Figure~\ref{f:ang} for a visualization of this).
The trough in the global $\beta$ profile arises because the dips in each quadrant occur at different radii and have different depths.

Interestingly, when we look at the stars that belonged to the satellite galaxy that merged with the system at $z\sim1$, these stars are evenly dispersed at all radii and angular cross-section.
However, when we look at the distribution of stars today that belonged to the progenitor of $g14\_h258$ at $z\sim1.4$, we see an overdensity of stars that looks like a tidal tail that wraps nearly around the galaxy.
When we look at $\beta$ in different angular quadrants, we pick out regions that cross this tidal structure. 
That is, the $\beta$ dip is, in fact, picking up stars that once belonged to the primary \textit{in situ} disk, but have been displaced in an extended tidal feature that enhances $\sigma_{\phi}$.
While visually this extended tidal feature is hard to disentangle from the overall stellar halo today, it has persisted from $z\sim0.8$ until the present, and the merger has left a lasting fingerprint on its kinematics.

While a merger event such as the one seen in \textit{g14\_h258} may rarely occur, the kinematic record should be long lasting, with $\beta\leq0$ over a wide range of radii at present day.
Hunting for a broad $\beta$ trough in the  global $\beta$ profile of MW could be of great value because it would give us deep insight into the MW's major merger history.
With the upcoming all-sky \textit{Gaia} survey and several follow up surveys to obtain line-of-sight velocities it will soon be possible to search for $\beta$ dips in many different parts of the sky and to use these observations to construct a global $\beta$ profile for the Galaxy.

\section{Discussion and Conclusions}
\label{s:conclusion}

The results and implications of this study are as follows:
\begin{enumerate}

\item{Both accretion-only simulations and $N$--body+SPH simulations predict strongly radially anisotropic velocity dispersions in the stellar halos for most MW-like disk galaxies.
The most robust observations in the Milky Way at $r=5-10$ kpc give $\beta=0.5-0.7$, which is consistent with predictions from simulations.}

\item{There are three situations under which low positive to negative values of $\beta$ arise in these MW-like simulations:}
  
\begin{enumerate}
\item{Transient passage and disruption of a satellite which contributes  a coherently moving group of stars to the stellar halo: such dips are short lived and last no longer than $\sim0.2$ Gyr.}

\item{Passage of a massive satellite  (that stays bound) through the inner part of the stellar halo induces transient changes in the kinematics of \textit{in situ} halo stars.
Dips in the \textit{in situ} halo are longer lived (lasting $> 0.2$ Gyr) and more metal-rich (on average $\sim 0.7$ dex higher) than dips in the accreted halo.}

\item{A major merger with another disk at high redshift ($z\sim1$) can generate a stellar halo with a $\beta$ trough -- significant tangential anisotropy over a range of radii -- which persists to the present day.
  Such a trough is likely to be comprised of multiple 10-30~kpc $\beta$ dips occurring at a range of radii which collectively appear as an extended trough.  These dips should be visible over a significant portion (at least one quarter to half) of the sky at any given radius.}
\end{enumerate}
\end{enumerate}
Previous results for $\beta$ at $r\sim20-30$ kpc in the MW based on proper-motions (measured by \textit{Hubble Space Telescope} in the direction of M31) suggest that $\beta$ could be nearly zero or even slightly negative \citep{Deason2013b, Cunningham2016}.
Such a low value of $\beta$ could arise from substructure (as has been proposed by \citealp{Deason2013b}).
If upcoming \textit{Gaia} data confirms this dip in $\beta$, we predict that, if it was produced by recently disrupted satellite, then the $\beta$ dip should be fairly localized in radius and unlikely to extend to over a larger portion of the sky. 
If this dip is found to be present primarily in higher metallicity stars than those typically found in the accreted stellar halo, it could point to the presence of an  \textit{in situ} stellar halo that was perturbed by the passage of a massive satellite.
In the unlikely event that the dip is found over a large portion of the sky and is highly negative over a wide range of radii, it could point to a major merger with a disk in the past.
Such a trough is likely to be comprised of multiple 10-30~kpc dips occurring at a range of radii.
These broad dips should be seen over a large portion of the sky, and the severity of a given dip is likely to differ in different parts of the sky.

It is clear that dips in $\beta$ in the MW are a sensitive probe of recent interactions with satellites and long ago interactions with other disk galaxies.
Determining proper-motions with \textit{Gaia} and fully characterizing 6D phase-space with future surveys like WFIRST \citep{WFIRST2015} will enable us to explore substructure in the stellar halo in a new way.
We posit that $\beta$ should be thought of as a tool for discovery, as it will enable us to find and follow-up on the building blocks of our stellar halo.

Finally, as mentioned in the introduction, one of the original motivations for determining the anisotropy parameter $\beta$ is that this quantity appears in the spherical form of the Jeans equations \citep{Jeans1915} and knowledge of $\beta(r)$ in the stellar halo would enable a determination of the mass profile of the MW's dark matter halo.
However the assumption underlying the use of the spherical Jeans equation is that the tracer population and the potential that it traces are relaxed (virialized) and in dynamical equilibrium.
As we have seen non-monotonic $\beta$ profiles generally arise from substructure or perturbations which are clear evidence for a halo out of dynamical equilibrium.
Since unvirialized systems tend to have higher kinetic energy than virialized systems the assumption of virial equilibrium would lead to an over-estimate in the halo mass.
Furthermore for a given  3D velocity dispersion, an inferred tangential anisotropy also results in a higher estimate of the dynamical mass.
This implies that if $\beta$ in the MW stellar halo is found to be negative 
due to its non-equilibrium state, then dynamical measurements of the halo mass that use $\beta$ are likely to overestimate the mass of the dark matter halo.

\section{Acknowledgments}
\label{s:ack}

We thank the anonymous referee for the useful feedback; the final manuscript is much stronger for their questions and comments.
S.R.L. also thanks Jillian Bellovary for the suggestion and support in exploring \textit{g14\_h258}.
S.R.L. acknowledges support from the Michigan Society of Fellows.
S.R.L. was also supported  by  NASA through Hubble Fellowship grant HST-HF2-51395.001-A from the Space  Telescope  Science  Institute,  which  is  operated  by  the Association of Universities for Research in Astronomy, Incorporated,  under  NASA  contract  NAS5-26555.
M.V. and K.H. are supported by NASA ATP award NNX15AK79G.
V.P.D. is supported by STFC Consolidated grant ST/M000877/1.
V.P.D. acknowledges being a part of the network supported by the COST Action TD1403 “Big Data Era in Sky and Earth Observation.” 
V.P.D. acknowledges the support of the Pauli Center for Theoretical Studies,
which is supported by the Swiss National Science Foundation (SNF), the
University of Z\"urich, and ETH Z\"urich and George Lake for
arranging for his sabbatical visit during which time this paper was
completed.
V.P.D. acknowledges the Michigan Institute of Research in Astrophysics (MIRA), which funded his collaboration visit to the University of Michigan during which research for this paper was completed.


\appendix
\label{s:appendix}

\section{Prospects for measuring a $\beta$-dip with \textit{Gaia} data}

In this appendix we estimate how accurately $\beta$ can be determined with \textit{Gaia} data, under a few simple assumptions, by analyzing mock catalogs of K giants and blue horizontal branch (BHB) stars with realistic observational errors.
We assume that the observational error on the line-of-sight velocity is small; this assumption is based on knowledge of current and future ground-based follow-up surveys, such as Gaia-ESO survey \citep{Gilmore2012}.
Gaia-ESO has attained line-of-sight velocity errors on order a few $\kms$; these errors are approximately valid for tracer populations such as BHB stars \citep{Xue2011} and K giants \citep{Xue2014}.  

We generate our mock catalogs assuming that the density profile of the stellar halo is given by $\rho(r) \propto r^{-3}$.
We assume that halo stars obey an anisotropic Gaussian velocity distribution specified by the velocity dispersions $(\sigma_r, \sigma_\phi, \sigma_\theta)$, and  that the system has no net rotation.
We also assume that these properties are independent of stellar type.
In addition, we assume that the radial velocity dispersion is independent of $r$ and is equal to $\sigma_r = 220 \kms / \sqrt{2} = 156 \kms$.   
We adopt two models for the tangential components of the velocity dispersion:
\eq{
\frac{\sigma_\phi^2(r)}{\sigma_r^2} = \frac{\sigma_\theta^2(r)}{\sigma_r^2} = 
\begin{cases}
E(r, 22.5 \kpc, 2.5 \kpc), \;\;(\mathrm{Model ~1}),\\
E(r, 22.5 \kpc, 5 \kpc), \;\;(\mathrm{Model ~2}),\\
\end{cases}
}
where we define 
\eq{
E(r, c, w) = 
\begin{cases}
\frac{3}{4} + \frac{1}{4} \cos [ \frac{2 \pi}{2 w} (r - c)] , \;\; (c-w < r < c+w), \\
\frac{1}{2}, \;\; (\mathrm{otherwise}) .
\end{cases}
}
Both Models 1 and 2 have a constant value of $\beta=0.5$ at $r<c-w$ and $c+w<r$, but $\beta(r)$ dips in between, attaining its minimum value of $\beta=0$ at $r=c=22.5 \kpc$.
The parameter $w$ determines the width of the low-$\beta$ region (dip), 
and the dip in Model 1 is sharp ($w=2.5 \kpc$)  while in Model 2  it is broad ($w=5 \kpc$). 

\begin{table}[h]
\caption{Assumed properties of mock stars}
\label{table:mock_catalog}
\centering
\begin{tabular}{ccc}
	\hline
	Sample & K giants & BHB stars\\
	\hline\hline
	$(V-I)$\footnote{(V-I) color} & $0.99$ mag & $0.5$ mag \\
	$M_V$\footnote{Absolute $V$ magnitude}  & $1.53$ mag & $0.71$ mag \\
	$\sigma_{DM}$\footnote{Error in distance modulus}  & $0.35$ mag & $0.10$ mag \\
	$\sigma_{v}$\footnote{Error in line-of-sight velocity}  & $5 \kms$ & $5 \kms$ \\
	\hline
\end{tabular}
\end{table}

For each model, we generate 2000 stars that satisfy 
$15 \kpc < r_\mathrm{obs} < 30 \kpc$, $|z_\mathrm{obs}|> 5 \kpc$, and $|b|> 30^\circ$. 
Here, $r_\mathrm{obs}$, $|z_\mathrm{obs}|$, and $b$ are 
the observed Galactocentric radius, vertical distance from the Galactic plane, 
and the Galactic latitude, respectively. 
The assumed distance modulus error ($\sigma_{DM}$) for K giants and BHB stars are shown in Table \ref{table:mock_catalog}. 
The line-of-sight velocity error is always assumed to be $\sigma_v = 5 \kms$. 
The assumed values of $(V-I, M_V)$ for K giants and BHB stars are shown in Table \ref{table:mock_catalog}, 
and these values are used to evaluate the end-of-mission \textit{Gaia} proper motion errors 
(with the publicly available code  \PyGaia \footnote{
{https://github.com/agabrown/PyGaia}}).

\begin{figure} 
\begin{center}
	\includegraphics[angle=0,width=0.445\columnwidth]{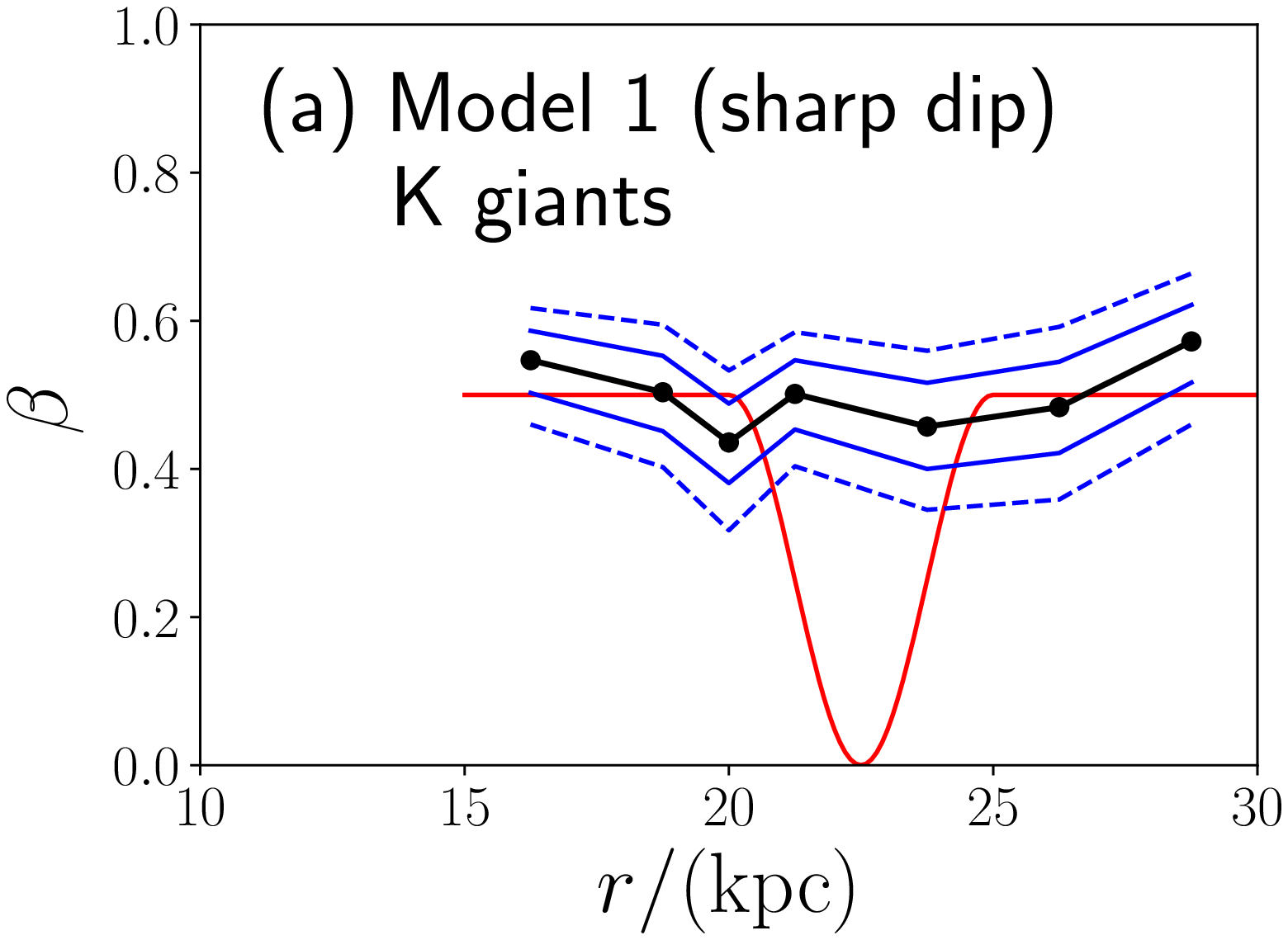} 
	\includegraphics[angle=0,width=0.445\columnwidth]{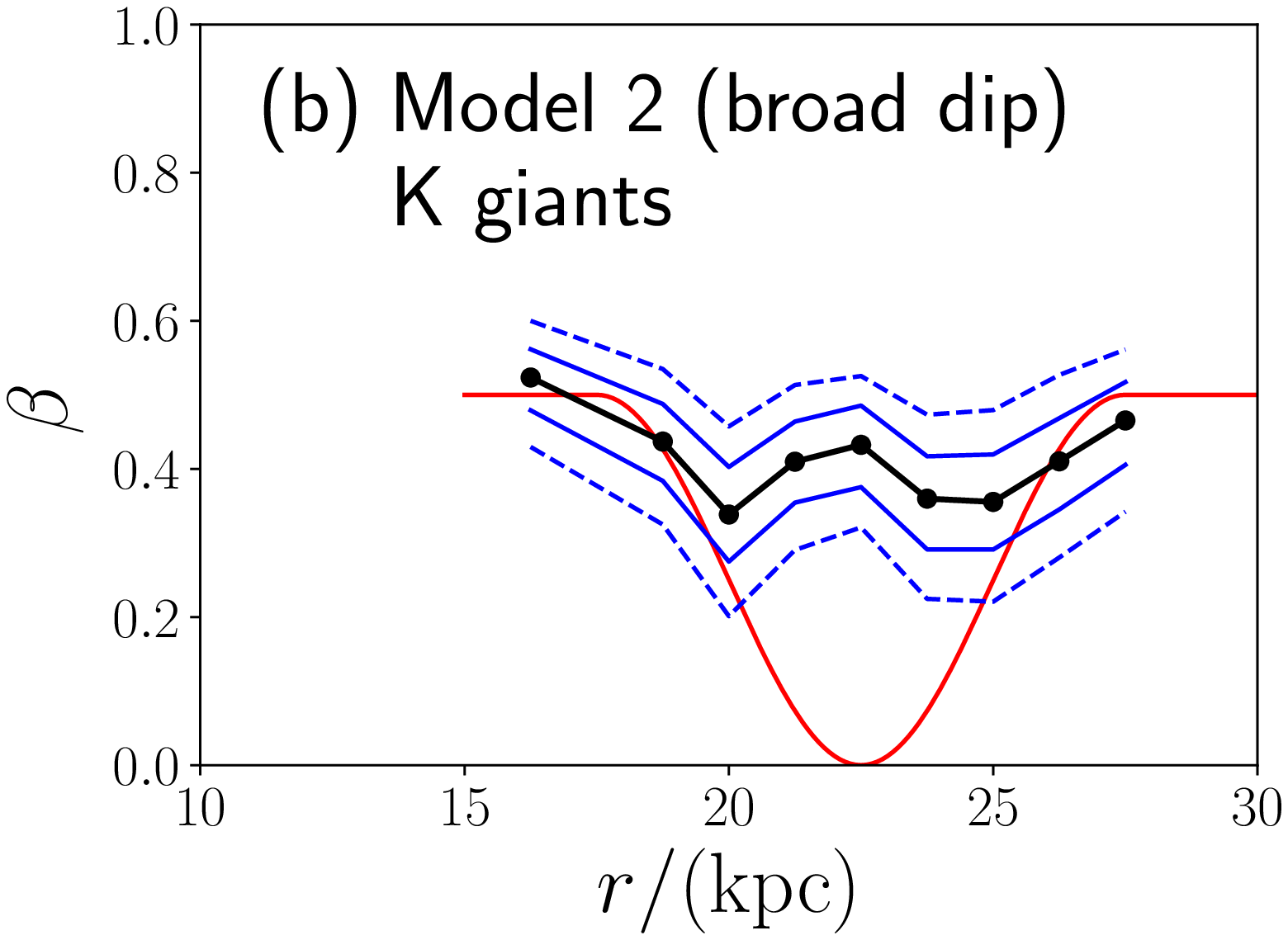} \\
	\includegraphics[angle=0,width=0.445\columnwidth]{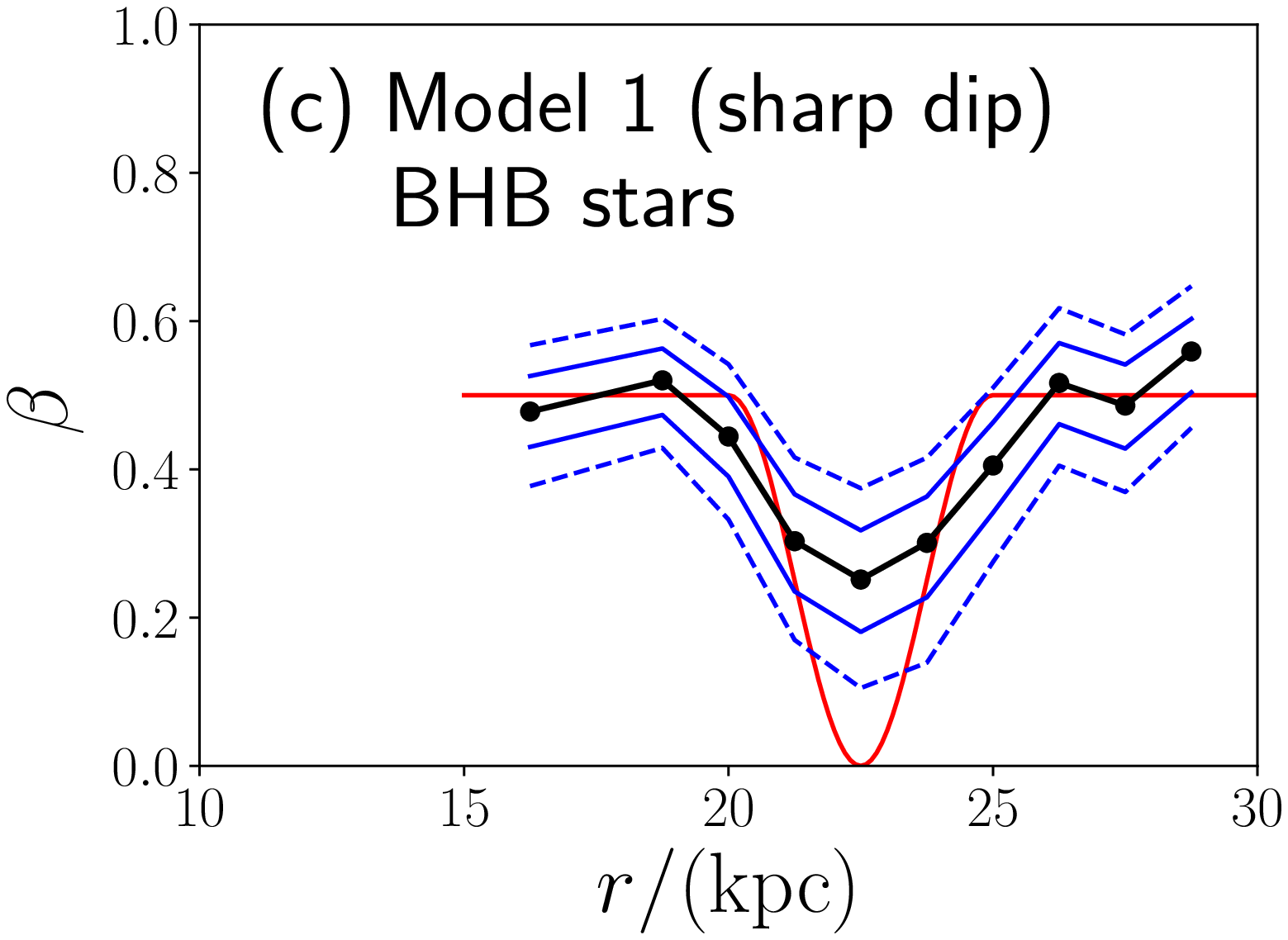} 
	\includegraphics[angle=0,width=0.445\columnwidth]{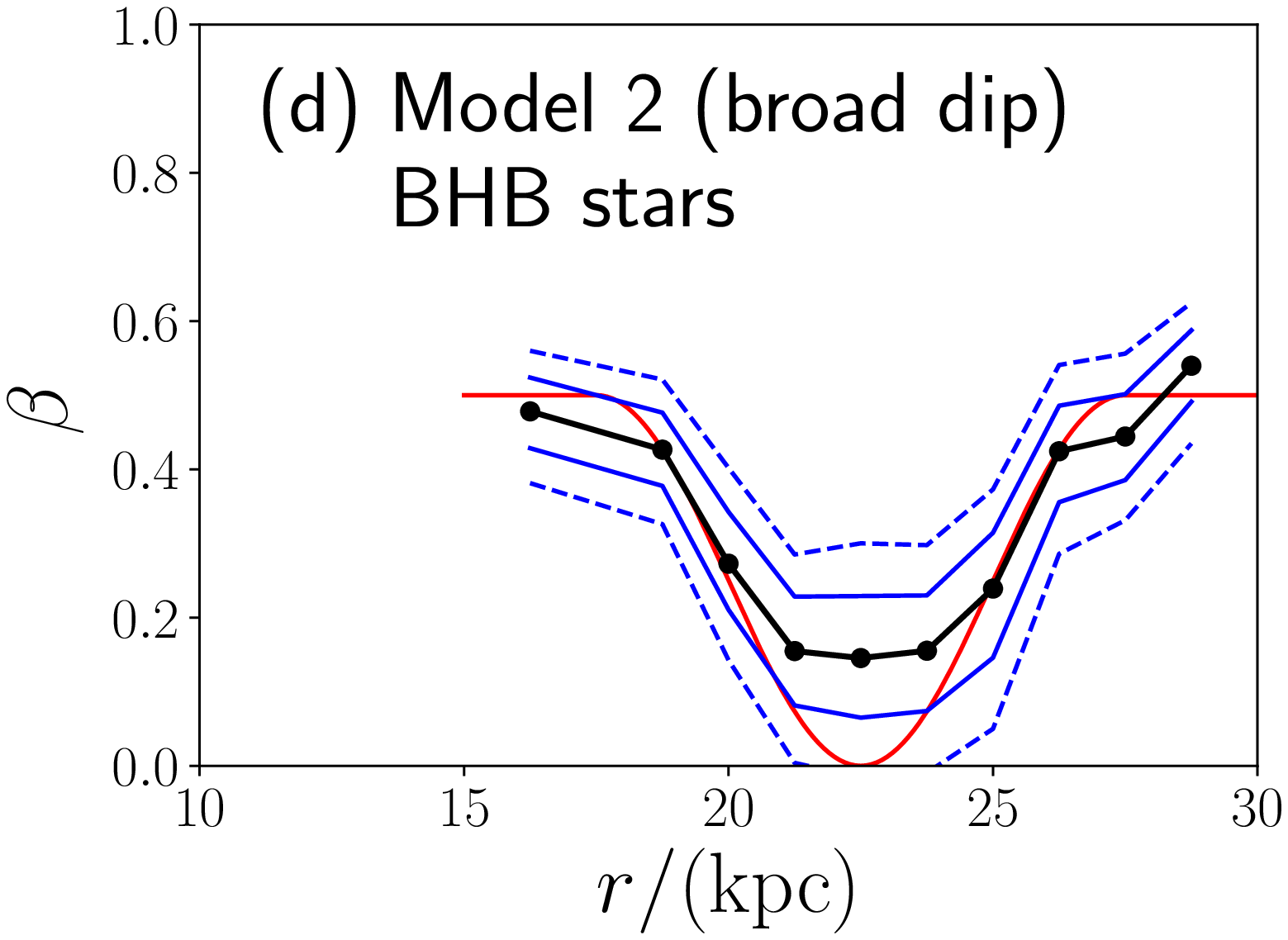} 
\end{center}
\caption{
Mock analyses of K giant and BHB star catalogs with \textit{Gaia}-like proper motion error. 
The input profile of $\beta(r)$ is shown with red curve. 
The black dot shows the the posterior median value of $\beta$ at each radial bin. 
Blue solid and dashed lines covers 68\% and 95\% of the posterior distribution of $\beta$, respectively.
We see that BHB stars are expected to be helpful in detecting dips in $\beta$ due to their small distance uncertainty (5\%), while K giants (distance error of 16\%) are not.
}
\label{fig:appendix}
\end{figure}

We generate two mock catalogs (Models 1 and 2) for each type of star (K giants and BHB stars). 
For each of these four mock catalogs, 
we performed Bayesian analysis (similar to that presented in \citealt{Hattori2017} and \citealt{Deason2017})
to derive the posterior distribution of $(\sigma_r, \sigma_\phi, \sigma_\theta)$. 
Figure \ref{fig:appendix} shows the recovered $\beta(r)$ profiles for each of our mock catalogs. 

The top two panels of this figure show that, for the mock K giant samples, neither the broad nor the narrow dip in the $\beta(r)$ profile can be recovered.
This is mainly because the distance error for K giants (16\%) is too large. 
For example, if the heliocentric distance of a K giant is $22.5 \kpc$, 
the associated distance error is $3.6 \kpc$, which is comparable to or larger than the radial extent of the dip, $w$, in our models. 
Thus, the sample stars with $r_\mathrm{obs}\simeq 22.5 \kpc$ 
are highly contaminated by foreground and background stars, 
so that the $\beta$ dip is blurred.  

When the mock BHB samples are used, the dips in the $\beta(r)$ profiles are recovered easily, although the depths are underestimated.
The dips in the BHB samples are more detectable than the dips in the K giant samples because the distance error for BHB stars (5\%) is small. 
In Model 2, the recovered $\beta$ profile is a very good match to the true $\beta$ profile at $15 \kpc < r < 30 \kpc$.
For both models, the  depth of the recovered profiles are underestimated, but the location of the dips near $r=22.5 \kpc$ are recovered quite accurately. 

These results suggest that in order to have the highest probability of detecting dips in the $\beta(r)$ profile with \textit{Gaia} proper motion data, we need to use halo tracers whose distance error is smaller than the radial extent of the dip.
Since the width (radial extent) of a dip is unknown {\it a priori}, it desirable to use a population for which the distance errors are small, like BHB stars.
Although BHB stars are less numerous than K giants, more than 2000 stars within 30~kpc have already been observed \citep{Xue2011}.
Since these BHB stars are brighter than the limiting magnitude of \textit{Gaia}, proper-motion will be obtained for all of them.
It is therefore likely that \textit{Gaia} data for BHB stars is capable of confirming the alleged dip in $\beta(r)$ profile at $r \simeq 20 \kpc$ (e.g., \citealt{Kafle2012, King2015}). 


\begin{thebibliography}{}

\bibitem[\protect\citeauthoryear{{Abadi}, {Navarro}, \& {Steinmetz}}{{Abadi}
  et~al.}{2006}]{Abadi2006}
{Abadi}, M.~G., {Navarro}, J.~F.,  \& {Steinmetz}, M. 2006, \mnras, 365, 747

\bibitem[\protect\citeauthoryear{{Besla} et~al.}{{Besla}
  et~al.}{2012}]{Besla2012}
{Besla}, G., {Kallivayalil}, N., {Hernquist}, L., {van der Marel}, R.~P.,
  {Cox}, T.~J.,  \& {Kere{\v s}}, D. 2012, \mnras, 421, 2109

\bibitem[\protect\citeauthoryear{{Bell} et~al.}{{Bell} et~al.}{2008}]{Bell2008}
{Bell}, E.~F., et~al. 2008, \apj, 680, 295

\bibitem[\protect\citeauthoryear{{Binney}}{{Binney}}{1980}]{Binney1980}
{Binney}, J. 1980, \mnras, 190, 873

\bibitem[\protect\citeauthoryear{{Binney} \& {Tremaine}}{{Binney} \&
  {Tremaine}}{2008}]{Binney2008}
{Binney}, J.,  \& {Tremaine}, S. 2008, {Galactic Dynamics: Second Edition}
  (Princeton University Press)

\bibitem[\protect\citeauthoryear{{Bird} \& {Flynn}}{{Bird} \&
  {Flynn}}{2015}]{Bird2015}
{Bird}, S.~A.,  \& {Flynn}, C. 2015, \mnras, 452, 2675

\bibitem[\protect\citeauthoryear{{Bond} et~al.}{{Bond} et~al.}{2010}]{Bond2010}
{Bond}, N.~A., et~al. 2010, \apj, 716, 1

\bibitem[\protect\citeauthoryear{{Brooks} \& {Zolotov}}{{Brooks} \&
  {Zolotov}}{2014}]{Brooks2014}
{Brooks}, A.~M.,  \& {Zolotov}, A. 2014, \apj, 786, 87

\bibitem[\protect\citeauthoryear{{Bullock} \& {Johnston}}{{Bullock} \&
  {Johnston}}{2005}]{Bullock2005}
{Bullock}, J.~S.,  \& {Johnston}, K.~V. 2005, \apj, 635, 931

\bibitem[\protect\citeauthoryear{{Chiba} \& {Yoshii}}{{Chiba} \&
  {Yoshii}}{1998}]{Chiba1998}
{Chiba}, M.,  \& {Yoshii}, Y. 1998, \aj, 115, 168

\bibitem[\protect\citeauthoryear{{Christensen} et~al.}{{Christensen}
  et~al.}{2012}]{Christensen2012}
{Christensen}, C., {Quinn}, T., {Governato}, F., {Stilp}, A., {Shen}, S.,  \&
  {Wadsley}, J. 2012, \mnras, 425, 3058

\bibitem[\protect\citeauthoryear{{Cunningham} et~al.}{{Cunningham}
  et~al.}{2015}]{Cunningham2015}
{Cunningham}, E.~C., {Deason}, A., {Guhathakurta}, P., {Rockosi}, C., {Kirby},
  E., {van der marel}, r.~p.,  \& {Sohn}, S.~T. 2015, IAU General Assembly, 22,
  2255864

\bibitem[\protect\citeauthoryear{{Cunningham} et~al.}{{Cunningham}
  et~al.}{2016}]{Cunningham2016}
{Cunningham}, E.~C., et~al. 2016, \apj, 820, 18

\bibitem[\protect\citeauthoryear{{Deason} et~al.}{{Deason}
  et~al.}{2012}]{Deason2012}
{Deason}, A.~J., {Belokurov}, V., {Evans}, N.~W.,  \& {An}, J. 2012, \mnras,
  424, L44

\bibitem[\protect\citeauthoryear{{Deason} et~al.}{{Deason}
  et~al.}{2013a}]{Deason2013a}
{Deason}, A.~J., {Belokurov}, V., {Evans}, N.~W.,  \& {Johnston}, K.~V. 2013a,
  \apj, 763, 113

\bibitem[\protect\citeauthoryear{{Deason} et~al.}{{Deason}
  et~al.}{2013b}]{Deason2013b}
{Deason}, A.~J., {Van der Marel}, R.~P., {Guhathakurta}, P., {Sohn}, S.~T.,  \&
  {Brown}, T.~M. 2013b, \apj, 766, 24
  
\bibitem[Deason et al.(2017)]{Deason2017} Deason, A.~J., Belokurov, V., Koposov, S.~E., et al.\ 2017, \mnras, 470, 1259 

\bibitem[\protect\citeauthoryear{{Debattista} et~al.}{{Debattista}
  et~al.}{2008}]{Debattista2008}
{Debattista}, V.~P., {Moore}, B., {Quinn}, T., {Kazantzidis}, S., {Maas}, R.,
  {Mayer}, L., {Read}, J.,  \& {Stadel}, J. 2008, \apj, 681, 1076

\bibitem[DESI Collaboration et al.(2016)]{DESI2016} DESI Collaboration, Aghamousa, A., Aguilar, J., et al.\ 2016, arXiv:1611.00036 
  
\bibitem[\protect\citeauthoryear{{Diemand}, {Madau}, \& {Moore}}{{Diemand}
  et~al.}{2005}]{Diemand2005}
{Diemand}, J., {Madau}, P.,  \& {Moore}, B. 2005, \mnras, 364, 367

\bibitem[\protect\citeauthoryear{{Eggen}, {Lynden-Bell}, \& {Sandage}}{{Eggen}
  et~al.}{1962}]{Eggen1962}
{Eggen}, O.~J., {Lynden-Bell}, D.,  \& {Sandage}, A.~R. 1962, \apj, 136, 748

\bibitem[\protect\citeauthoryear{{Font} et~al.}{{Font} et~al.}{2006}]{Font2006}
{Font}, A.~S., {Johnston}, K.~V., {Bullock}, J.~S.,  \& {Robertson}, B.~E.
  2006, \apj, 638, 585

\bibitem[\protect\citeauthoryear{{Gaia Collaboration} et~al.}{{Gaia
  Collaboration} et~al.}{2016}]{Gaia2016}
{Gaia Collaboration}, et~al. 2016, \aap, 595, A1

\bibitem[\protect\citeauthoryear{{Gill}, {Knebe}, \& {Gibson}}{{Gill}
  et~al.}{2004}]{AHF2004}
{Gill}, S.~P.~D., {Knebe}, A.,  \& {Gibson}, B.~K. 2004, \mnras, 351, 399

\bibitem[\protect\citeauthoryear{{Gilmore} et~al.}{{Gilmore} et~al.}{2012}]{Gilmore2012} {Gilmore}, G., et~al. 2012, The Messenger, 147, 25
  
\bibitem[\protect\citeauthoryear{{Gnedin} et~al.}{{Gnedin}
  et~al.}{2010}]{Gnedin2010}
{Gnedin}, O.~Y., {Brown}, W.~R., {Geller}, M.~J.,  \& {Kenyon}, S.~J. 2010,
\apjl, 720, L108

\bibitem[\protect\citeauthoryear{{Governato} et~al.}{{Governato}
  et~al.}{2009}]{Governato2009}
{Governato}, F., et~al. 2009, \mnras, 398, 312

\bibitem[\protect\citeauthoryear{{Governato} et~al.}{{Governato}
  et~al.}{2012}]{Governato2012}
{Governato}, F., et~al. 2012, \mnras, 422, 1231

\bibitem[\protect\citeauthoryear{{Hattori} et~al.}{{Hattori}
  et~al.}{2013}]{Hattori2013}
{Hattori}, K., {Yoshii}, Y., {Beers}, T.~C., {Carollo}, D.,  \& {Lee}, Y.~S.
  2013, \apjl, 763, L17

\bibitem[Hattori et al.(2017)]{Hattori2017} Hattori, K., Valluri, M., Loebman, S.~R., \& Bell, E.~F.\ 2017, \apj, 841, 91 


\bibitem[\protect\citeauthoryear{{Jeans}}{{Jeans}}{1915}]{Jeans1915}
{Jeans}, J.~H. 1915, \mnras, 76, 70

\bibitem[\protect\citeauthoryear{{Johnston} et~al.}{{Johnston}
  et~al.}{2008}]{Johnston2008}
{Johnston}, K.~V., {Bullock}, J.~S., {Sharma}, S., {Font}, A., {Robertson},
  B.~E.,  \& {Leitner}, S.~N. 2008, \apj, 689, 936

\bibitem[\protect\citeauthoryear{{Kafle} et~al.}{{Kafle}
  et~al.}{2012}]{Kafle2012}
{Kafle}, P.~R., {Sharma}, S., {Lewis}, G.~F.,  \& {Bland-Hawthorn}, J. 2012,
  \apj, 761, 98

\bibitem[\protect\citeauthoryear{{Katz} \& {White}}{{Katz} \&
  {White}}{1993}]{Katz1993}
{Katz}, N.,  \& {White}, S.~D.~M. 1993, \apj, 412, 455

\bibitem[\protect\citeauthoryear{{King} et~al.}{{King} et~al.}{2015}]{King2015}
{King}, C., III, {Brown}, W.~R., {Geller}, M.~J.,  \& {Kenyon}, S.~J. 2015,
  \apj, 813, 89

\bibitem[\protect\citeauthoryear{{Knollmann} \& {Knebe}}{{Knollmann} \&
  {Knebe}}{2009}]{AHF2009}
{Knollmann}, S.~R.,  \& {Knebe}, A. 2009, \apjs, 182, 608

\bibitem[\protect\citeauthoryear{{Munn} et~al.}{{Munn} et~al.}{2004}]{Munn2004}
{Munn}, J.~A., et~al. 2004, \aj, 127, 3034

\bibitem[\protect\citeauthoryear{{Munshi} et~al.}{{Munshi}
  et~al.}{2013}]{Munshi2013}
{Munshi}, F., et~al. 2013, \apj, 766, 56

\bibitem[\protect\citeauthoryear{{Pillepich}, {Madau}, \& {Mayer}}{{Pillepich}
  et~al.}{2015}]{Pillepich15}
{Pillepich}, A., {Madau}, P.,  \& {Mayer}, L. 2015, \apj, 799, 184

\bibitem[\protect\citeauthoryear{{Rashkov} et~al.}{{Rashkov}
  et~al.}{2013}]{Rashov2013}
{Rashkov}, V., {Pillepich}, A., {Deason}, A.~J., {Madau}, P., {Rockosi}, C.~M.,
  {Guedes}, J.,  \& {Mayer}, L. 2013, \apjl, 773, L32

\bibitem[\protect\citeauthoryear{{Robertson} et~al.}{{Robertson}
  et~al.}{2005}]{Robertson2005}
{Robertson}, B., {Bullock}, J.~S., {Font}, A.~S., {Johnston}, K.~V.,  \&
  {Hernquist}, L. 2005, \apj, 632, 872

\bibitem[\protect\citeauthoryear{{Sales} et~al.}{{Sales}
  et~al.}{2007}]{Sales2007}
{Sales}, L.~V., {Navarro}, J.~F., {Abadi}, M.~G.,  \& {Steinmetz}, M. 2007,
\mnras, 379, 1464

\bibitem[\protect\citeauthoryear{{Santos-Santos} et~al.}{{Santos-Santos}
  et~al.}{2016}]{Santos-Santos2016}
{Santos-Santos}, I.~M., {Brook}, C.~B., {Stinson}, G., {Di Cintio}, A.,
  {Wadsley}, J., {Dom{\'{\i}}nguez-Tenreiro}, R., {Gottl{\"o}ber}, S.,  \&
  {Yepes}, G. 2016, \mnras, 455, 476

\bibitem[\protect\citeauthoryear{{Shen}, {Wadsley}, \& {Stinson}}{{Shen}
  et~al.}{2010}]{Shen2010}
{Shen}, S., {Wadsley}, J.,  \& {Stinson}, G. 2010, \mnras, 407, 1581

\bibitem[\protect\citeauthoryear{{Sirko} et~al.}{{Sirko}
  et~al.}{2004}]{Sirko2004}
{Sirko}, E., et~al. 2004, \aj, 127, 914

\bibitem[\protect\citeauthoryear{{Smith} et~al.}{{Smith}
  et~al.}{2009}]{Smith2009}
{Smith}, M.~C., et~al. 2009, \mnras, 399, 1223

\bibitem[\protect\citeauthoryear{{Snaith} et~al.}{{Snaith}
  et~al.}{2016}]{Snaith2016}
{Snaith}, O.~N., {Bailin}, J., {Gibson}, B.~K., {Bell}, E.~F., {Stinson}, G.,
  {Valluri}, M., {Wadsley}, J.,  \& {Couchman}, H. 2016, \mnras, 456, 3119

\bibitem[\protect\citeauthoryear{{Somerville} \& {Kolatt}}{{Somerville} \&
  {Kolatt}}{1999}]{Somerville1999}
{Somerville}, R.~S.,  \& {Kolatt}, T.~S. 1999, \mnras, 305, 1

\bibitem[\protect\citeauthoryear{{Spergel} et~al.}{{Spergel}
  et~al.}{2015}]{WFIRST2015}
{Spergel}, D., et~al. 2015, ArXiv e-prints

\bibitem[\protect\citeauthoryear{{Spergel} et~al.}{{Spergel}
  et~al.}{2003}]{Spergel2003}
{Spergel}, D.~N., et~al. 2003, \apjs, 148, 175

\bibitem[\protect\citeauthoryear{Stinson et~al.}{Stinson
  et~al.}{2013}]{Stinson2012}
Stinson, G., Brook, C., Maccio, A.~V., Wadsley, J., Quinn, T.~R.,  \& Couchman,
  H. M.~P. 2013, Mon. Not. Roy. Astron. Soc., 428, 129

\bibitem[\protect\citeauthoryear{{Stinson} et~al.}{{Stinson}
  et~al.}{2006}]{Stinson2006}
{Stinson}, G., {Seth}, A., {Katz}, N., {Wadsley}, J., {Governato}, F.,  \&
  {Quinn}, T. 2006, \mnras, 373, 1074

\bibitem[\protect\citeauthoryear{{Tissera} et~al.}{{Tissera}
  et~al.}{2013}]{Tissera2013}
{Tissera}, P.~B., {Scannapieco}, C., {Beers}, T.~C.,  \& {Carollo}, D. 2013,
  \mnras, 432, 3391

\bibitem[\protect\citeauthoryear{{Valluri} et~al.}{{Valluri}
  et~al.}{2016}]{Valluri2016}
{Valluri}, M., {Loebman}, S.~R., {Bailin}, J., {Clarke}, A., {Debattista},
  V.~P.,  \& {Stinson}, G. 2016, in IAU Symposium, Vol. 317, The General
  Assembly of Galaxy Halos: Structure, Origin and Evolution, ed.
  A.~{Bragaglia}, M.~{Arnaboldi}, M.~{Rejkuba}, \& D.~{Romano}, 358

\bibitem[\protect\citeauthoryear{{Wadsley}, {Stadel}, \& {Quinn}}{{Wadsley}
  et~al.}{2004}]{Wadsley2004}
{Wadsley}, J.~W., {Stadel}, J.,  \& {Quinn}, T. 2004, New Astronomy, 9, 137

\bibitem[\protect\citeauthoryear{{Wang} et~al.}{{Wang} et~al.}{2015}]{Wang2015}
{Wang}, W., {Han}, J., {Cooper}, A.~P., {Cole}, S., {Frenk}, C.,  \& {Lowing},
  B. 2015, \mnras, 453, 377

\bibitem[\protect\citeauthoryear{{Wilkinson} \& {Evans}}{{Wilkinson} \&
  {Evans}}{1999}]{Wilkinson1999}
{Wilkinson}, M.~I.,  \& {Evans}, N.~W. 1999, \mnras, 310, 645

\bibitem[\protect\citeauthoryear{{Williams} \& {Evans}}{{Williams} \&
  {Evans}}{2015}]{Williams2015}
{Williams}, A.~A.,  \& {Evans}, N.~W. 2015, \mnras, 454, 698

\bibitem[\protect\citeauthoryear{{Xue} et~al.}{{Xue} et~al.}{2014}]{Xue2014}{Xue}, X.-X., et~al.\ 2014, \apj, 784, 170
  
\bibitem[\protect\citeauthoryear{{Xue} et~al.}{{Xue} et~al.}{2011}]{Xue2011}
{Xue}, X.-X., et~al. 2011, \apj, 738, 79

\bibitem[\protect\citeauthoryear{{Xue} et~al.}{{Xue} et~al.}{2008}]{Xue2008}
  {Xue}, X.~X., et~al. 2008, \apj, 684, 1143

\bibitem[\protect\citeauthoryear{{Zolotov} et~al.}{{Zolotov}
  et~al.}{2012}]{Zolotov2012}
{Zolotov}, A., et~al. 2012, \apj, 761, 71

\bibitem[\protect\citeauthoryear{{Zolotov} et~al.}{{Zolotov}
  et~al.}{2009}]{Zolotov2009}
{Zolotov}, A., {Willman}, B., {Brooks}, A.~M., {Governato}, F., {Brook}, C.~B.,
  {Hogg}, D.~W., {Quinn}, T.,  \& {Stinson}, G. 2009, \apj, 702, 1058
  
\end{thebibliography}
\end{document}